\documentclass[fleqn,usenatbib]{mnras}

\usepackage{newtxtext,newtxmath}

\usepackage[T1]{fontenc}
\usepackage{ae,aecompl}
\usepackage{multirow}

\usepackage{graphicx}	
\usepackage{amsmath}	
\usepackage{amssymb}	

\newcommand{\lya}{Ly$\alpha$ }
\title[Ly$\alpha$ window into cosmic reionization]{Ly$\alpha$ forest power spectrum as an emerging window into the epoch of reionization and cosmic dawn}

\author[Montero-Camacho and Mao]{
Paulo Montero-Camacho,\thanks{pmontero@tsinghua.edu.cn (PMC)}
and Yi Mao\thanks{ymao@tsinghua.edu.cn (YM)} 
\\
Department of Astronomy, Tsinghua University, Beijing 100084, China\\
}

\date{Accepted 2020 September 18. Received 2020 September 18; in original form 2020 March 27}

\pubyear{2020}

\begin{document}
\label{firstpage}
\pagerange{\pageref{firstpage}--\pageref{lastpage}}
\maketitle

\begin{abstract}
Conventional wisdom was that thermal relics from the epoch of reionization (EOR) would vanish swiftly. Recently, however, it was shown that these relics can survive to lower redshifts ($z \sim 2$) than previously thought, due to gas at mean density being heated to $T \sim 3 \times 10^4$ K by reionization, which is inhomogeneous, and shocks. Given the high sensitivities of upcoming \lya forest surveys, this effect will be a novel broadband systematic for cosmological application. From the astrophysical point of view, however, the imprint of inhomogeneous reionization can shed light on the EOR and cosmic dawn. 
We utilize a hybrid method --- which includes two different simulation codes capable of handling the huge dynamical range --- to show the impact of patchy reionization on the \lya forest and its dependence on different astrophysical scenarios. We found statistically significant deviations in the 1D \lya power spectrum at $k = 0.14$ cMpc$^{-1}$ that range from $\sim 1\%$ at $z = 2$ up to almost $\sim 20\%$ at $z = 4$. The deviations in the 3D \lya power spectrum, at the same wavenumber, are large and range from a few per cent at $z = 2$ up to $\sim 50\%$ at $z = 4$, although these deviations ignore the effect of \ion{He}{II} reionization and AGN feedback at $z<4$. By exploiting different $k$-dependence of power spectrum among various astrophysical scenarios, the effect of patchy reionization on the \lya forest power spectrum can open a new window into cosmic reionization and possibly cosmic dawn. 
\end{abstract}

\begin{keywords}
methods: numerical  --- galaxies: intergalactic medium --- cosmology: dark ages, reionization, first stars 
\end{keywords}



\section{Introduction}
\label{sec:intro}
After the surface of last scattering ($z_{\rm dec} \sim 1059$) the gas in the Universe became transparent to the cosmic microwave background (CMB) photons. As the Universe expanded and cooled, eventually complex structures, such as stars and galaxies, formed thanks to gravitational instabilities. These objects emit ultraviolet (UV) photons, which ultimately reionize the Universe and heat up the intergalactic medium (IGM) to $\sim 10^4 \, \textup{K}$ \citep[see, e.g.,][]{2016ARA&A..54..313M, 2019ApJ...874..154D}. After cosmic reionization, which is currently believed to occur halfway around $z_{\rm re} = 7.68$ \citep{2018arXiv180706209P}, the absorption features of neutral hydrogen regions in quasar spectra, i.e. the \lya forest, stand as one of the primary probes of the IGM at the redshifts $2<z<6$. 

Among other important probes, the \lya forest has been used to investigate \ion{H} I reionization, particularly its end at $z\sim 6$ \citep[see, e.g.,][]{2002AJ....123.1247F, 2009ApJ...706L.164C, 2009ApJ...694..842M, 2015MNRAS.447..499M}. The study of the effect of hydrogen reionization in the \lya forest has traditionally focused on the highest redshifts at $z\sim 5-6$ \citep[e.g.][]{1997ApJ...486..599H, 2008ApJ...689L..81T, 2014ApJ...788..175L, 2016MNRAS.463.2335N}, where the IGM has not yet relaxed into the usual temperature-density relation \citep{2009ApJ...701...94F} and there are enough sightlines for robust statistics. Moreover, the \lya forest has also been used to study \ion{He}{II} reionization -- which is believed to occur around $z \sim 3.5$  \citep[see, e.g.,][]{2014MNRAS.445.4186C, 2016ApJ...825..144W, 2019ApJ...875..111W, 2017ApJ...838...96K}. 

At redshifts lower than the end of \ion{H} I reionization, say $2<z<4$, conventional wisdom was that thermal relics from the epoch of reionization (EOR) would vanish swiftly and hence the usual IGM temperature-density relation would be recovered rapidly. As such, \lya forest can probe the cosmological large-scale structure at the post-reionization epoch. However, the sensitivity of the forest to the high redshift IGM can possibly lead to new interesting challenges for current and future \lya forest surveys. Recent works have shown that the impact of inhomogeneous reionization in the \lya forest is large at high redshifts \citep{2019MNRAS.487.1047M, 2019MNRAS.490.3177W,  2019MNRAS.486.4075O}, and can survive to lower redshifts ($z \sim 2$) than previously thought and is even comparable to instrumental sensitivities at lower redshifts \citep{2019MNRAS.487.1047M}. This novel low-redshift result is primarily due to high-entropy mean-density (HEMD) gas that is heated to $T \sim 3 \times 10^4\,{\rm K}$ \citep{2018MNRAS.474.2173H} by reionization and subsequent shocks from higher density regions, coupled to the inhomogeneous nature of \ion{H}{I} reionization. Note that when we refer to the imprint of inhomogeneous reionization throughout this paper, it is really the imprint of early structure formation coupled with the patchy nature of reionization. In order to resolve this HEMD phase of the temperature-density evolution, we utilize high-resolution hydrodynamic simulations capable of resolving the neutral gas to below the Jeans mass prior to reionization. Also, we include streaming velocities between baryon and dark matter since they suppress the amount of small-scale structure.
The effect of inhomogeneous reionization -- with the HEMD gas -- on the \lya forest power spectrum, as a novel broadband systematic, imposes a challenge to the \lya forest for its promise to do precision cosmology at higher redshifts $z\sim 3-4$. For this purpose, substantial efforts must be made to transform these first studies of the effect of patchy reionization in the \lya forest into holistic precision cosmology programs. 

Furthermore, recent developments in the \lya forest (including large scale fluctuations in its opacity and damping wing studies) and \lya emission have thrown the status of the redshift of reionization into an open question \citep{2015MNRAS.447..499M,2015MNRAS.447.3402B, 2018MNRAS.479.1055B, 2018ApJ...864...53E, 2019ApJ...878...12H, 2019MNRAS.489.2669M, 2019arXiv191205582K}. Interestingly, a possible emerging consensus points to a later reionization than inferred from the \textit{Planck}'s optical depth. This is relevant in the context of the imprint of inhomogeneous reionization in the \lya forest. In particular, if islands of neutral hydrogen are indeed floating around at $z < 6$ \citep{2019MNRAS.485L..24K, 2019arXiv191003570N, 2020MNRAS.491.1736K}, one should expect a stronger impact than that computed in \cite{2019MNRAS.487.1047M} for the later reionization model (their model A). 
In tandem to the theoretical and computational recent discussions, the upcoming observational efforts are coming online. For example, the Dark Energy Spectroscopic Instrument \citep[DESI;][]{2016arXiv161100036D} will soon start to measure a plethora of \lya skewers and begin its \lya science program.     

Although the scenario might be challenging, the \lya forest is not alone. The 21~cm hyperfine transition of hydrogen will ultimately supplement the \lya forest as yet another rich probe of the EOR and cosmic dawn. \cite{2018Natur.555...67B} reported a likely first measurement of the global signal of the 21~cm brightness temperature. Besides, the 21~cm global signal has already been used to rule out some sudden reionization scenarios \citep{2010Natur.468..796B, 2017ApJ...847...64M, 2018ApJ...858...54S}, and to study the astrophysics of high redshift hydrogen gas \citep{2018ApJ...863...11M}. Furthermore, as pointed out in \cite{2019MNRAS.487.1047M}, in principle the quadrupole of the 21~cm power spectrum can be used to mitigate the effect of patchy reionization in the \lya forest. For further details in the anticipated fruits of the 21~cm revolution, see, e.g., \cite{10.1088/2514-3433/ab4a73}. 

From the astrophysical point of view, on the other hand, the imprint of inhomogeneous reionization in the \lya forest power spectrum can shed light on the EOR and cosmic dawn. 
This paper has two main goals. First, we shall illustrate how the effect of patchy reionization on the \lya forest power spectrum, even at low redshifts, can become a window into the EOR and cosmic dawn. Secondly, we shall explore the dependence of this effect within the astrophysical parameter space, and investigate different $k$-dependence of power spectrum which may be exploited to distinguish various astrophysical scenarios. Such a study can help build the connection between the 21~cm cosmology and \lya forest. 

This paper is organized as follows. We outline our simulation strategy for handling the huge dynamical range involved with the effect of inhomogeneous reionization in the \lya forest and for obtaining the necessary ingredients for our calculations in \S\ref{sec:methods}. We describe the key astrophysical parameters used to model the reionization process that were allowed to vary and the different models constructed from them in \S\ref{sec:models}. In \S\ref{sec:relic}, we report the impact of inhomogeneous reionization in the \lya forest, both for the 1D and 3D power spectra, for all our models. 
We summarize our results and discuss future work in \S\ref{sec:conclu}.


\section{Methodology}
\label{sec:methods}

\subsection{Simulations}
In order to compute the effect of patchy reionization in the \lya forest, the small-scale physics must be resolved to simulate the behavior of gas, while large box simulations are required to capture the inhomogeneous nature of reionization. Here we follow the approach by \cite{2019MNRAS.486.4075O, 2019MNRAS.487.1047M} to overcome these obstacles. We split the tasks since the dynamical range is too large with only one simulation. We use the modified {\sc Gadget2} code \citep{2005MNRAS.364.1105S} from \cite{2018MNRAS.474.2173H} to resolve the gas to below the Jeans mass prior to reionization. These small-box simulations have sudden reionization and do not include any prescription to add ionizing sources. From the small-scale simulations, we obtain an optical depth map of how the transmission of the IGM depends on when reionization happens.
Meanwhile, we use the {\tt 21cmFASTv1.3} code \citep{2007ApJ...669..663M,2011MNRAS.411..955M} with minor modifications from \cite{2019MNRAS.487.1047M} to tackle the patchy nature of reionization on the large scale. We chose this version for simple comparisons with our previous work. The large-scale simulations can extract the effect of patchiness on the matter distribution, specifically the cross-power spectrum of matter and neutral hydrogen fraction, i.e. how matter and bubble spatial structure are correlated. With these ingredients we can calculate the effect of inhomogeneous reionization on the \lya forest power spectrum.

For small-scale simulations, we describe the key physical ingredients present here, but refer interested readers to \cite{2018MNRAS.474.2173H} for a full description of our small box simulations, including tests, convergence and details on how to construct a \lya forest from such small boxes. The simulations used here correspond to the II-F simulations from \cite{2018MNRAS.474.2173H}, which have a box size of $2551 \ \textup{ckpc}$ on each side with the particle number of $2 \times (384)^2$. Furthermore, the dark matter particle mass is $9.72 \times 10^3 \ \textup{M}_\odot$ and the gas mass is $1.81 \times 10^3 \ \textup{M}_\odot$. We have implemented streaming velocities between baryons and dark matter \citep{2010PhRvD..82h3520T, 2020arXiv200212296G}, which modulate the amount of small-scale structure (if baryons are moving faster they might not fall into a specific potential well). In each simulation, reionization happens suddenly at one of the following redshifts: 6, 7, 8, 9, 10, 11 or 12. We ran eight different realizations for each simulation in order to reduce the variance in the inferred transparency of the IGM. The small-scale simulations evolve the neutral gas since recombination up to cosmic reionization. Reionization is implemented by immediately changing the temperature of all particles to $2 \times 10^4 \ \textup{K}$. After reionization the simulation evolves the particles by singly ionized primordial gas physics for H$^{+}$ and He$^{+}$.   

There is no \ion{He}{II} cosmic reionization implemented in the small-scale simulations. We note that this is problematic since the effects of inhomogeneous reionization in the \lya forest have not relaxed into the usual temperature-density relation by $z \sim 3$ \citep{2019MNRAS.487.1047M}.

For large-scale simulations, we use a simulation box size $400$~Mpc on each side, larger than the box size of 300 Mpc used in \cite{2019MNRAS.487.1047M}, with $256^3$ ($768^3$) cells for \ion{H} I (matter) field. Furthermore, we run four different realizations for each reionization scenario with {\tt 21cmFAST} in order to reduce the variance in our simulations. As a result, we can compute the sample variance on the mean of our models, and therefore obtain the error associated with the cross-power spectrum.

We summarize the multiple limitations of our approach in \S\ref{ssec:limi}.

Throughout this work, we use the cosmological parameters from the full \textit{Planck} 2015 release \citep{2016A&A...594A..13P}, given by $\Omega_{\rm m}h^2 = 0.14170$, $\Omega_{\rm b}h^2 = 0.02230$, $\sigma_8 = 0.8159$, $n_s = 0.9667$ and $H_0 = 67.74 \ \textup{km} \, \textup{s}^{-1} \, \textup{Mpc}^{-1}$. 

\subsection{Ly$\alpha$ forest power spectrum}
The formalism for the effect of patchy reionization on the \lya forest power spectrum is described in detail in \S2 of \cite{2019MNRAS.487.1047M}. Here we recapitulate the important points of the derivation. 

The fluctuations on the transmitted \lya flux under the effect of patchy reionization can be written as \footnote{
Throughout this paper, our convention is that fluctuations are defined as  $\delta_p = p/\bar{p} - 1$ for any observable except for the neutral hydrogen fraction, where the fluctuation is given by $\delta_{x_{\rm HI}} = x_{\rm HI} - \bar{x}_{\rm HI}$. Moreover, the auto-power spectrum of $\delta_p$ is written as $P_p$, and the cross-power spectrum of $\delta_p$ with $\delta_{p'}$ is written as $P_{p,p'}$. Besides, all power spectra are in dimensionless form, $k^3 P(k) / 2\pi^2$.}
\begin{eqnarray}
    \label{eq:lya}
    \delta_{\rm F}(\boldsymbol{k}, z_{\rm obs} ) = (1 + \beta_{\rm F}\mu^2)\, b_{\rm F} \, \delta_m (\boldsymbol{k}, z_{\rm obs}) + b_{\rm \Gamma} \, \psi(\boldsymbol{k}, z_{\rm obs}, z_{\rm re}) \, \textup{,}
\end{eqnarray}
where we have explicitly shown the dependence of the different fluctuations involved on the redshift $z_{\rm obs}$ and wavenumber $\boldsymbol{k}$. Here $b_{\rm F}$ is the usual flux bias parameter, $\beta_{\rm F}$ the redshift-space distortion (RSD) parameter, and $b_{\rm \Gamma} = \partial \ln \bar{F} / \partial \ln \tau_1$ is the radiation bias parameter. Its role here is to convert optical depth changes into flux fluctuations. $\tau_1$ corresponds to the optical depth needed for the mean flux from a patch of gas with temperature $10^4 \ \textup{K}$ and density $\Delta_b = 1$ to reproduce the observed transmitted flux. Because we vary the normalization $\tau_1$ to match the observed flux in our small-scale simulations, the change to the \lya forest due to reionization happening suddenly at redshift $z_{\rm re}$ in comparison to redshift $\bar{z}_{\rm re}$ is reported as a change in transparency, and optical depth, of the IGM, which we parametrize as $ \Delta \ln \tau_1 (z_{\rm re}, \bar{z}_{\rm re}) = \ln [\tau_1(z_{\rm re}) / \tau_1 (\bar{z}_{\rm re})] = \psi (z_{\rm re})$ for each observed redshift $z_{\rm obs}$. The results of small-scale simulations are used to compute the transmission $ \psi (z_{\rm re},z_{\rm obs})$. Since our simulations are the same  as in \cite{2019MNRAS.487.1047M}, the function of $\psi (z_{\rm re},z_{\rm obs})$ takes the values listed in their Table 3. 

The 3D power spectrum of the transmitted flux of the \lya forest, ignoring higher order terms in $\psi$ and computed perpendicular to the line of sight, is given by
\begin{eqnarray}
\label{eq:p3D}
P^{\rm 3D}_{\rm F} (k, z_{\rm obs}) \simeq b_{\rm F}^2 P_m (k, z_{\rm obs}) + 2b_{\rm F}b_{\rm \Gamma}P_{m,\psi} (k, z_{\rm obs}) \, \textup{,} 
\end{eqnarray}
where
\begin{eqnarray}
\label{eq:psicross}
P_{m, \psi}(k, z_{\rm obs}) = - \int_{z_{\rm min}}^{z_{\rm max}} dz \frac{\partial \psi}{\partial z}(z,z_{\rm obs}) P_{m,x_{\rm HI}}(k,z) \frac{D(z_{\rm obs})}{D(z)} \, \textup{.}
\end{eqnarray}
Here we set the lower limit of integration $z_{\rm min} = 5.90$ and the upper limit $z_{\rm max} = 34.7$. 
Even though we set the integration limits to cover most of the reionization history, the peak of the contribution to the integral roughly comes from when the Universe is half reionized. The integral has three different contributions: growth rate ratio, $D(z_{\rm obs})/D(z)$ which is only affected by cosmology, the change of transparency of the IGM, $\partial \psi/\partial z\,(z,z_{\rm obs})$ computed by assuming sudden reionization scenarios in our small-scale high-resolution simulations, and inhomogeneous reionization $P_{m,x_{\rm HI}}$, which we vary for different reionization scenarios in this work. Interested readers are referred to \cite{2019MNRAS.487.1047M} for more details regarding the derivation and limitations of Eq.~(\ref{eq:psicross}). The cross-power spectrum of matter and transmission $P_{m,\psi}(k,z)$ is computed from the results of the large-scale reionization simulations, which are used to compute the cross-power spectrum of matter and the neutral fraction $P_{m,x_{\rm HI}}(k,z)$, together with the results of small-scale simulations, which are used to compute the transmission $\psi (z_{\rm re},z_{\rm obs})$ and its derivative. 

Throughout this work we utilize the same bias parameters used in \cite{2019MNRAS.487.1047M}, as summarized in their Table 2. Their radiation bias coefficients were obtained using their simulations and the flux bias coefficients were obtained from \cite{2011MNRAS.415.2257M}. Also, we follow their choice of setting $\beta_{\rm F} = 1$ in the matter-dominated Universe for consistency. 

In order to compute the 1D power spectrum of matter and transparency of the IGM, we integrate the second term of Eq.~(\ref{eq:p3D}) --- with the factor of $(1 + \mu^2)$ --- over the perpendicular direction. The methodology of converting the map from 3D to 1D is described explicitly in \S4.4 of \cite{2019MNRAS.487.1047M}. We directly extract the 1D \lya forest power spectrum from the data and then compare with the effect of patchy reionization. In order to estimate the errors due to simulations present herein, we have followed the same procedure described in \cite{2019MNRAS.487.1047M} with the only difference being the smaller amount of the reionization realizations (with the {\tt 21cmFAST} code) made in this work (four realizations herein compared to eight realizations in the previous work). The main budget of error comes from the small-scale simulations and therefore it is well-justified. Also, we have ignored the error from eBOSS since it is negligible in comparison to the variance in our simulations.

\begin{table*}
    \centering
    \caption{Different models explored throughout this work. The symbol ``---'' herein means that this parameter takes the fiducial value. Here $z_{\rm re}$ stands for the redshift when the Universe is halfway ionized. The values of optical depth are taken approximately for some of the models (T1, T2, and $\zeta 1$ models) wherein the reionization is not completely finished at the end of the large-scale simulations $z_{\rm min} = 5.90$.}
    \begin{tabular}{ccccccccc}
    \hline \hline
 {} &   Model  & $T_{\rm min}$ [K] & $R_{\rm mfp}$ [Mpc] & $\zeta$ & $E_0$ [eV] & $\zeta_{X}\,[\textup{M}_\odot^{-1}]$  & $z_{\rm re}$ & $\tau$  \\
    \hline
 {} &   fiducial  & $3 \times 10^4 $ & 50 & 25 & 500  & $2\times 10^{56}$ &  7.69  & 0.0547 \\
    \hline
\multirow{7}{*}{\rotatebox[origin=c]{90}{Bubble models}} &    T1        & $5 \times 10^4 $ & --- & --- & --- & ---   & 6.97  & 0.0477 \\
{} &    T2        & $4 \times 10^4 $ & --- & --- & --- & ---     & 7.28  & 0.0507 \\
{} &    T3        & $2 \times 10^4 $ & --- & --- & --- & ---     & 8.28  & 0.0607 \\
{} &    R1       & --- & 25 & --- & --- & ---    & 7.68  & 0.0543 \\
{} &    R2       & --- & 15 & --- & --- & ---   & 7.63  & 0.0536 \\
{} &    $\zeta_1$   & --- & --- & 20 & --- & ---  & 7.21  & 0.0504 \\
{} &    $\zeta_2$   & --- & --- & 30 & --- & ---& 8.08  & 0.0583  \\
    \hline
\multirow{6}{*}{\rotatebox[origin=c]{90}{Heating models}} &    E01        & --- & --- & --- & 100 & --- & 7.68  & 0.0552 \\
{} &    E02        & --- & --- & --- & 1000 & --- & 7.67  & 0.0545 \\
{} &    E03        & --- & --- & --- & 1500 & --- & 7.66  & 0.0544 \\
{} &    $\zeta_X1$     & --- & --- & --- & --- & $1 \times 10^{56}$    & 7.67  & 0.0546 \\
{} &    $\zeta_X$2     & --- & --- & --- & --- & $4 \times 10^{56}$    & 7.72  & 0.0550 \\
{} &    $\zeta_X$3     & --- & --- & --- & --- & $8 \times 10^{56}$    & 7.77  & 0.0556 \\
    \hline \hline 
    \end{tabular}
    \label{tab:models}
\end{table*}

\begin{table*}
\centering
\caption{Percentage deviation of the 3D \lya power spectrum due to patchy reionization, i.e.\ $2  (b_{\Gamma}/b_{\rm F}) P_{\rm m, \psi}/P_{\rm m} \times 100\% $ at $k = 0.14 \ \textup{Mpc}^{-1}$ (a typical scale for \lya surveys), at various redshifts for the different reionization and thermal models considered herein. The corresponding percentage deviation of the 1D \lya power spectrum due to patchy reionization is also shown in the lower part of this Table. 
For the 1D \lya power spectrum we have used the latest data release by BOSS + eBOSS \citep{2019JCAP...07..017C}. For the T1, T2, and $\zeta 1$ models with $z_{\rm re} < 7.30$ wherein reionization is not completely finished at the end of the large-scale simulations $z_{\rm min} = 5.90$, the change here represents only a lower limit. The errors in our results correspond to the sample variance in our simulations and are mainly dominated by the variance in our small-scale simulations.}
\label{tab:result}
\begin{tabular}{cccccccc}
\hline\hline
{} & {} & Model &  $z_{\rm obs} = 2.0$ & $z_{\rm obs} = 2.5$ & $z_{\rm obs} = 3.0$ & $z_{\rm obs} = 3.5$ & $z_{\rm obs} = 4.0$ \\
\hline
\multirow{16}{*}{\rotatebox[origin=c]{90}{3D Power Spectrum}} & {} & Fiducial  & $(4.46 \pm 0.55)\%$ & $(5.64 \pm 1.11)\%$ & $(10.0 \pm 1.69)\%$ & $(21.3 \pm 2.26)\%$ & $(38.2 \pm 3.06)\%$ \\
\cline{2-8}
{} & \multirow{7}{*}{\rotatebox[origin=c]{90}{Bubble models}} & T1  & $(5.81 \pm 0.62)\%$ & $(8.04 \pm 1.28)\%$ & $(14.1 \pm 1.86)\%$ & $(27.8 \pm 2.54)\%$ & $(47.5 \pm 3.61)\%$ \\
{} & {} & T2  & $(5.30 \pm 0.60)\%$ & $(7.11 \pm 1.22)\%$ & $(12.5 \pm 1.80)\%$ & $(25.4 \pm 2.43)\%$ & $(44.2 \pm 3.38)\%$ \\
{} & {} & T3  & $(3.19 \pm 0.48)\%$ & $(3.40 \pm 0.94)\%$ & $(6.14 \pm 1.48)\%$ & $(14.8 \pm 1.95)\%$ & $(28.5 \pm 2.50)\%$ \\
{} & {} & R1  & $(5.05 \pm 0.60)\%$ & $(6.54 \pm 1.21)\%$ & $(11.5 \pm 1.82)\%$ & $(24.0 \pm 2.43)\%$ & $(42.5 \pm 3.32)\%$ \\
{} & {} & R2  & $(6.24 \pm 0.69)\%$ & $(8.36 \pm 1.42)\%$ & $(14.6 \pm 2.09)\%$ & $(29.5 \pm 2.82)\%$ & $(51.3 \pm 3.93)\%$\\
{} & {} & $\zeta_1$  & $(5.53 \pm 0.61)\%$ & $(7.49 \pm 1.25)\%$ & $(13.1 \pm 1.83)\%$ & $(26.2 \pm 2.49)\%$ & $(45.3 \pm 3.50)\%$ \\
{} & {} & $\zeta_2$  & $(3.34 \pm 0.49)\%$ & $(3.71 \pm 0.96)\%$ & $(6.75 \pm 1.50)\%$ & $(15.9 \pm 1.98)\%$ & $(30.1 \pm 2.57)\%$ \\
\cline{2-8}
{} & \multirow{6}{*}{\rotatebox[origin=c]{90}{Heating models}} & E01  		  & $(3.88 \pm 0.51)\%$ & $(4.73 \pm 1.01)\%$ & $(8.50 \pm 1.56)\%$ & $(18.6 \pm 2.07)\%$ & $(34.0 \pm 2.75)\%$ \\
{} & {} & E02  		  & $(4.58 \pm 0.56)\%$ & $(5.82 \pm 1.13)\%$ & $(10.3 \pm 1.72)\%$ & $(21.9 \pm 2.29)\%$& $(39.1 \pm 3.12)\%$ \\
{} & {} & E03  		  & $(4.61 \pm 0.56)\%$ & $(5.87 \pm 1.13)\%$ & $(10.4 \pm 1.72)\%$ & $(22.0 \pm 2.31)\%$ & $(39.3 \pm 3.13)\%$ \\
{} & {} & $\zeta_X$1  & $(4.54 \pm 0.56)\%$ & $(5.76 \pm 1.12)\%$ & $(10.2 \pm 1.71)\%$ & $(21.7 \pm 2.28)\%$ & $(38.8 \pm 3.10)\%$ \\
{} & {} & $\zeta_X$2  & $(4.32 \pm 0.54)\%$ & $(5.42 \pm 1.08)\%$ & $(9.65 \pm 1.66)\%$ & $(20.7 \pm 2.21)\%$ & $(37.2 \pm 2.98)\%$ \\
{} & {} & $\zeta_X$3  & $(4.10 \pm 0.52)\%$ & $(5.07 \pm 1.05)\%$ & $(9.05 \pm 1.60)\%$ & $(19.6 \pm 2.14)\%$  & $(35.5 \pm 2.87)\%$ \\
\hline
\multirow{16}{*}{\rotatebox[origin=c]{90}{1D Power Spectrum}}  & {} & Fiducial  & $(1.07 \pm 0.13)\%$ & $(1.58 \pm 0.38)\%$ & $(3.12 \pm 0.66)\%$ & $(7.16 \pm 0.88)\%$ & $(14.8 \pm 1.25)\%$ \\
\cline{2-8}
{} & \multirow{7}{*}{\rotatebox[origin=c]{90}{Bubble models}} & T1  		  & $(1.40 \pm 0.14)\%$ & $(2.39 \pm 0.43)\%$ & $(4.68 \pm 0.71)\%$ & $(9.64 \pm 0.96)\%$ & $(18.8 \pm 1.44)\%$ \\
{} &  {} & T2  		  & $(1.28 \pm 0.14)\%$ & $(2.09 \pm 0.42)\%$ & $(4.10 \pm 0.70)\%$ & $(8.76 \pm 0.93)\%$ & $(17.4 \pm 1.38)\%$ \\
{} &  {} & T3  		  & $(0.78 \pm 0.12)\%$ & $(0.88 \pm 0.34)\%$ & $(1.73 \pm 0.60)\%$ & $(4.83 \pm 0.79)\%$ & $(11.0 \pm 1.07)\%$ \\
{} &  {} & R1  		  & $(1.16 \pm 0.14)\%$ & $(1.77 \pm 0.40)\%$ & $(3.47 \pm 0.69)\%$ & $(7.79 \pm 0.92)\%$ & $(15.9 \pm 1.32)\%$ \\
{} &  {} & R2  		  & $(1.40 \pm 0.15)\%$ & $(2.26 \pm 0.46)\%$ & $(4.40 \pm 0.78)\%$ & $(9.46 \pm 1.04)\%$ & $(18.9 \pm 1.53)\%$ \\
{} &  {} & $\zeta_1$  & $(1.39 \pm 0.15)\%$ & $(2.30 \pm 0.44)\%$ & $(4.49 \pm 0.74)\%$ & $(9.44 \pm 0.99)\%$ & $(18.6 \pm 1.47)\%$ \\
{} &  {} & $\zeta_2$    & $(0.79 \pm 0.12)\%$ & $(0.96 \pm 0.33)\%$ & $(1.92 \pm 0.59)\%$ & $(5.09 \pm 0.77)\%$ & $(11.3 \pm 1.06)\%$ \\
\cline{2-8}
{} & \multirow{6}{*}{\rotatebox[origin=c]{90}{Heating models}} & E01  		  & $(0.95 \pm 0.13)\%$ & $(1.34 \pm 0.36)\%$ & $(2.67 \pm 0.63)\%$ & $(6.38 \pm 0.82)\%$ & $(13.5 \pm 1.16)\%$ \\
{} &  {} & E02  		  & $(1.09 \pm 0.13)\%$ & $(1.63 \pm 0.39)\%$ & $(3.22 \pm 0.67)\%$ & $(7.34 \pm 0.89)\%$ & $(15.2 \pm 1.27)\%$ \\
{} &  {} & E03  		  & $(1.10 \pm 0.14)\%$ & $(1.65 \pm 0.39)\%$ & $(3.25 \pm 0.67)\%$ & $(7.40 \pm 0.89)\%$ & $(15.3 \pm 1.28)\%$ \\
{} &  {} & $\zeta_X$1  & $(1.08 \pm 0.13)\%$ & $(1.62 \pm 0.39)\%$ & $(3.19 \pm 0.67)\%$ & $(7.29 \pm 0.88)\%$ & $(15.1 \pm 1.27)\%$ \\
{} &  {} & $\zeta_X$2  & $(1.03 \pm 0.13)\%$ & $(1.52 \pm 0.38)\%$ & $(2.99 \pm 0.65)\%$ & $(6.93 \pm 0.86)\%$ & $(14.4 \pm 1.22)\%$ \\
{} &  {} & $\zeta_X$3  & $(0.98 \pm 0.13)\%$ & $(1.41 \pm 0.36)\%$ & $(2.79 \pm 0.63)\%$ & $(6.55 \pm 0.83)\%$ & $(13.8 \pm 1.18)\%$ \\
\hline\hline
\end{tabular}
\end{table*}

\section{Models of Reionization and Cosmic Dawn}
\label{sec:models}
In this paper we allow the variations of five astrophysical parameters used in the {\tt 21cmFAST} code, as follows. 

(1) $T_{\rm min}$, the minimum virial temperature of haloes that host ionizing sources. For haloes with virial temperatures smaller than this threshold, there is effectively no star formation in them. This temperature threshold plays a role in modulating the sources of ionizing photons and directly affects the properties of the reionization bubbles. If $T_{\rm min}$ increases (and all other astrophysical parameters are kept fixed), then less haloes of a given mass can have star-forming galaxies, which implies less UV photons available to ionize the Universe.

(2) $R_{\rm mfp}$, the mean free path of ionizing photons. It dictates the maximum horizon of ionizing photons, and defines the maximum permitted size of the bubbles, and hence a decrease in $R_{\rm mfp}$ implies more bubbles are needed to percolate the Universe, therefore one should expect a slight delay in the reionization process. 

(3) $\zeta$, the ionizing efficiency, i.e., roughly speaking, the number of ionizing photons that can escape from the stars into the IGM per each baryon atom in haloes. This parameter governs the timing of the reionization process in the {\tt 21cmFAST} code. The ionizing efficiency governs the amount of available UV photons that can ionize the \ion{H}{I} regions. If $\zeta$ increases, then there will be more UV photons in the IGM, thus reionization will happen sooner. 

(4) $E_0$, the energy threshold for the lowest energy X-ray photons not absorbed by galaxies. This parameter mainly affects the heating of the IGM prior to reionization. A larger value corresponds to inefficient X-ray heating of the IGM due to more X-ray photons being absorbed by the host galaxies, i.e. less photons preheat the IGM, and hence there is a slight delay in the reionization process.\footnote{Careful readers may find in Table~\ref{tab:models} that $z_{\rm re}$ for the E01 model is very slightly smaller than that for the fiducial model, which seems to contradict with the general trend here. However, the difference  $\Delta z_{\rm re} = 0.01$ between these two models is so small that it is actually due to numerical fluctuations of the different realizations. The comparison of $\tau$ between these two models is indeed consistent with the trend.}  

(5) $\zeta_{X}$, the X-ray efficiency which corresponds to the number of X-ray photons that manage to escape the galaxy per solar mass present in stars. The role of this parameter is to establish the preheating of the IGM. This X-ray efficiency controls the degree of X-ray heating that happens prior to reionization. Higher values would eventually cause reionization to occur earlier.

We chose these parameters inspired by the exploration of the impact of astrophysical parameters on the global 21~cm signal \citep{2018ApJ...863...11M}, and the effects on both the 21~cm fluctuations and the neutral hydrogen fraction (\citealt{2017MNRAS.472.2651G}; see their Figure 1). 
For the purpose of comparison, we use a fiducial model: $T_{\rm min} = 3 \times 10^4 \ \textup{K}$, $R_{\rm mfp} = 50 \ \textup{Mpc}$, $\zeta = 25$, $E_0 = 500 \ \textup{eV}$, and $\zeta_X = 2\times 10^{56} \ \textup{M}_\odot^{-1}$ (which corresponds to roughly 0.3 X-ray photons per stellar baryon). The reionization history in our fiducial model reproduces the optical depth of the \textit{Planck} result \citep{2018arXiv180706209P} quite well. 


We list all astrophysical models in Table \ref{tab:models}. For clarity, we group different models studied herein into two categories --- ``bubble models'', and ``heating models'' --- based on the primary role of the parameter allowed to vary. Specifically, the bubble models are those by varying three parameters, $T_{\rm min}$, $R_{\rm mfp}$ and $\zeta$, because their variations directly affect the growth or evolution of the ionized bubbles. On the other hand, the heating models correspond to the variations in $E_0$ and $\zeta_X$, because their variations mainly affect the preheating of the IGM. 
In Table \ref{tab:models}, we also list the redshifts of their halfway ionized epoch, $z_{\rm re}$, and the CMB optical depths $\tau$ corresponding to their global reionization histories. Even though the chosen models have variations of $z_{\rm re}$ of 1.11, we note that they are all loosely consistent with observational constraints and upper limits (see, e.g., Figure 12 of \citealt{2018ApJ...856....2M}).

\section{Results and Discussions}
\label{sec:relic}

\begin{figure}
    \centering
    \includegraphics[width=\columnwidth,keepaspectratio]{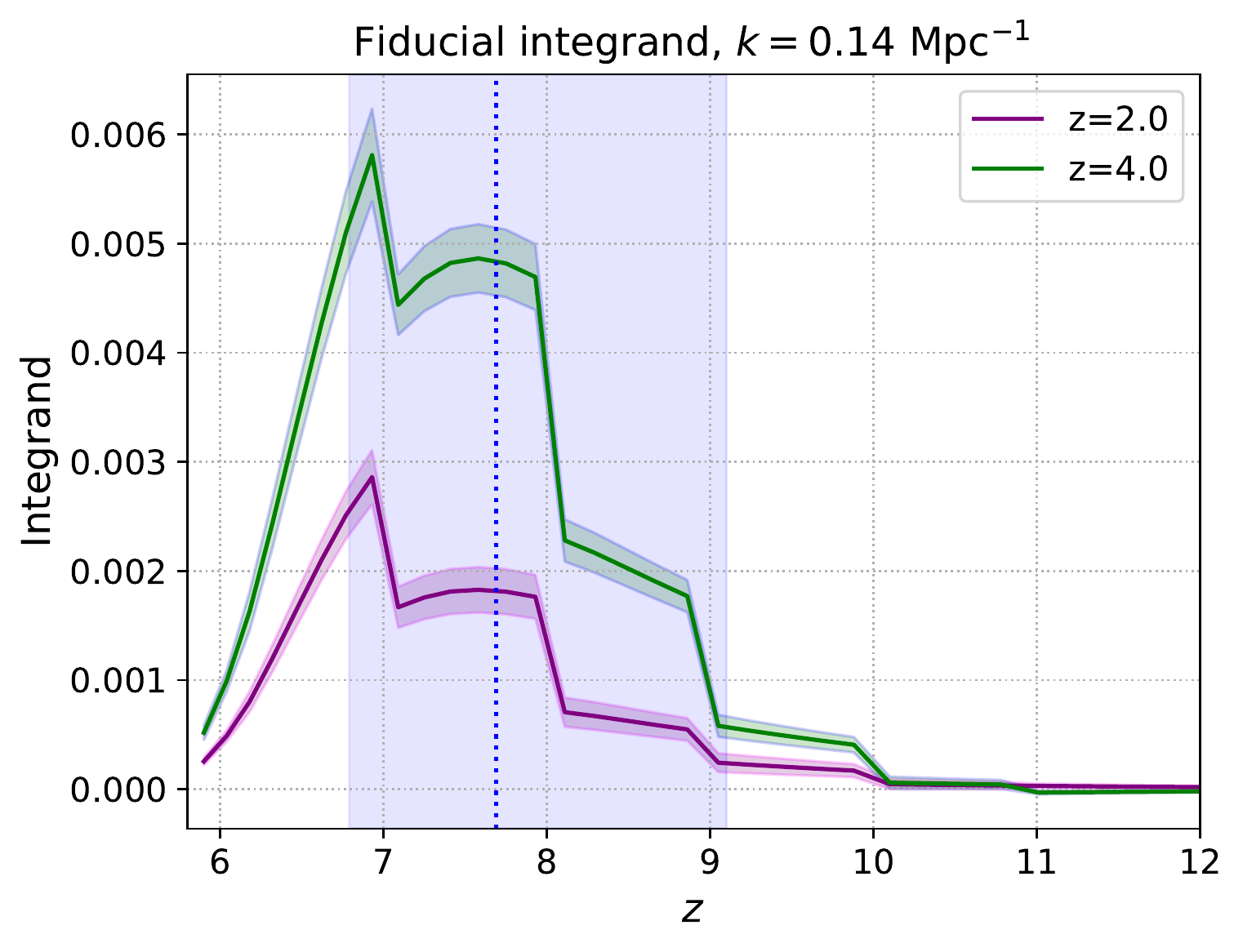}
    \caption{Integrand in Eq.~(\ref{eq:psicross}) as a function of redshift for low ($z_{\rm obs}=2$, purple) and high ($z_{\rm obs}=4$, green) redshift of observation for our fiducial model, with the shaded regions corresponding to the sample variance of our simulations. The dotted blue vertical line corresponds to the midpoint of reionization ($z_{\rm re}=7.69$), and the blue shaded region spans from when $\bar{x}_{\rm HI} = 0.25$
    ($z = 6.79$) up to $\bar{x}_{\rm HI} = 0.75$ ($z = 9.10$). We restrict the x-axis to show the peak of the contribution. The appreciable bumps in the integrand are due to the few redshifts of (sudden) reionization explored in our small-scale simulations. }
    \label{fig:fid_int}
\end{figure}

\subsection{The impact of inhomogeneous reionization}

\begin{figure}
    \centering
    \includegraphics[width=\columnwidth,keepaspectratio]{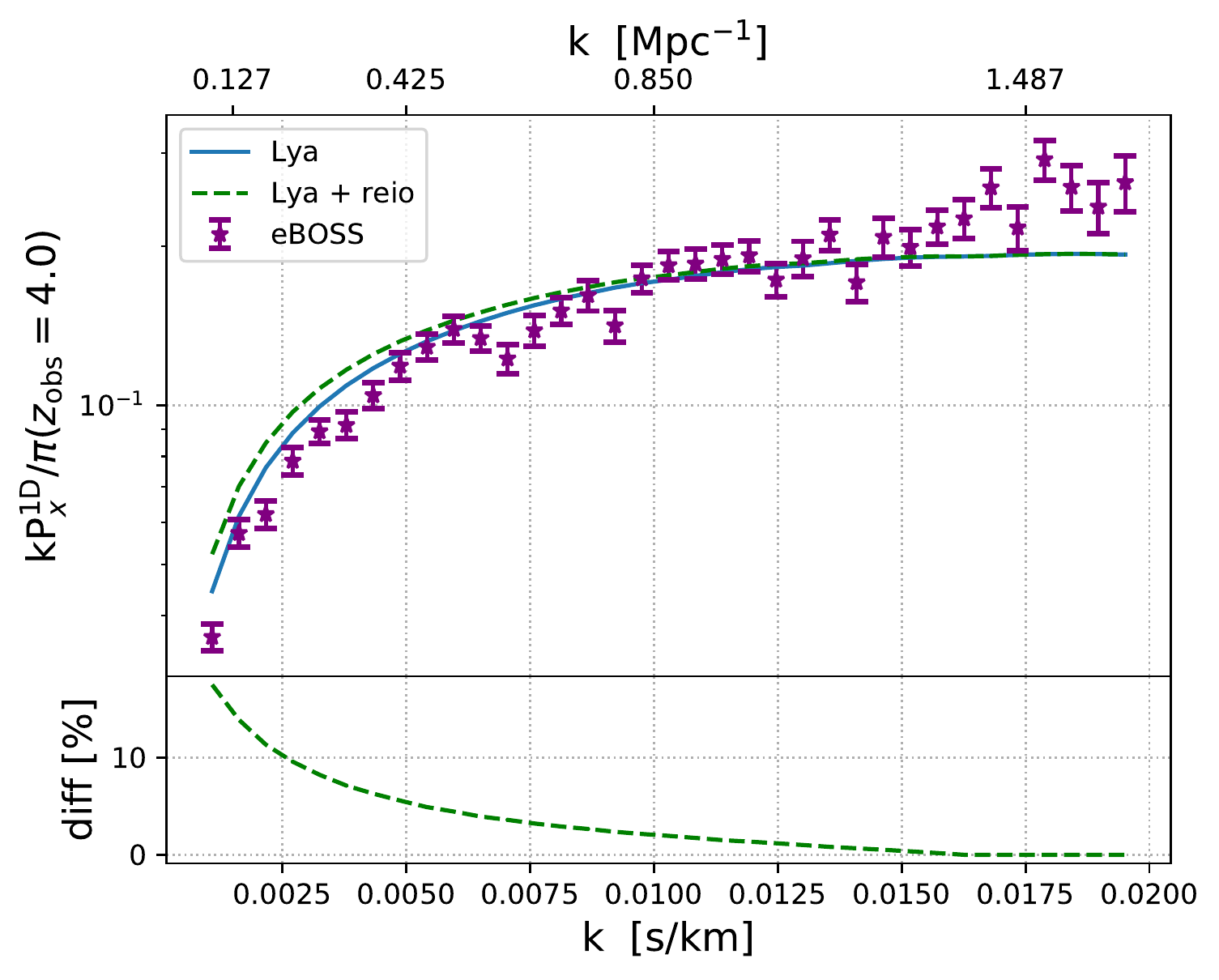}
    \caption{Memory of patchy reionization in the 1D \lya power spectrum for our fiducial model and observed at $z = 4.0$. (Top) we show the ``traditional'' \lya term of the power spectrum (i.e. $b_{\rm F}^2 P^{\rm 1D}_{m}$, blue solid line), the total power spectrum that includes the usual \lya term and the contribution from the memory of inhomogeneous reionization (green dashed line), and the data points from \citet{2019JCAP...07..017C} (purple points). (Bottom) percentage difference between the total 1D \lya power spectrum that includes the reionization effect and the traditional \lya power spectrum. }
    \label{fig:fid_1D}
\end{figure}

\begin{figure}
    \centering
    \includegraphics[width=\columnwidth,keepaspectratio]{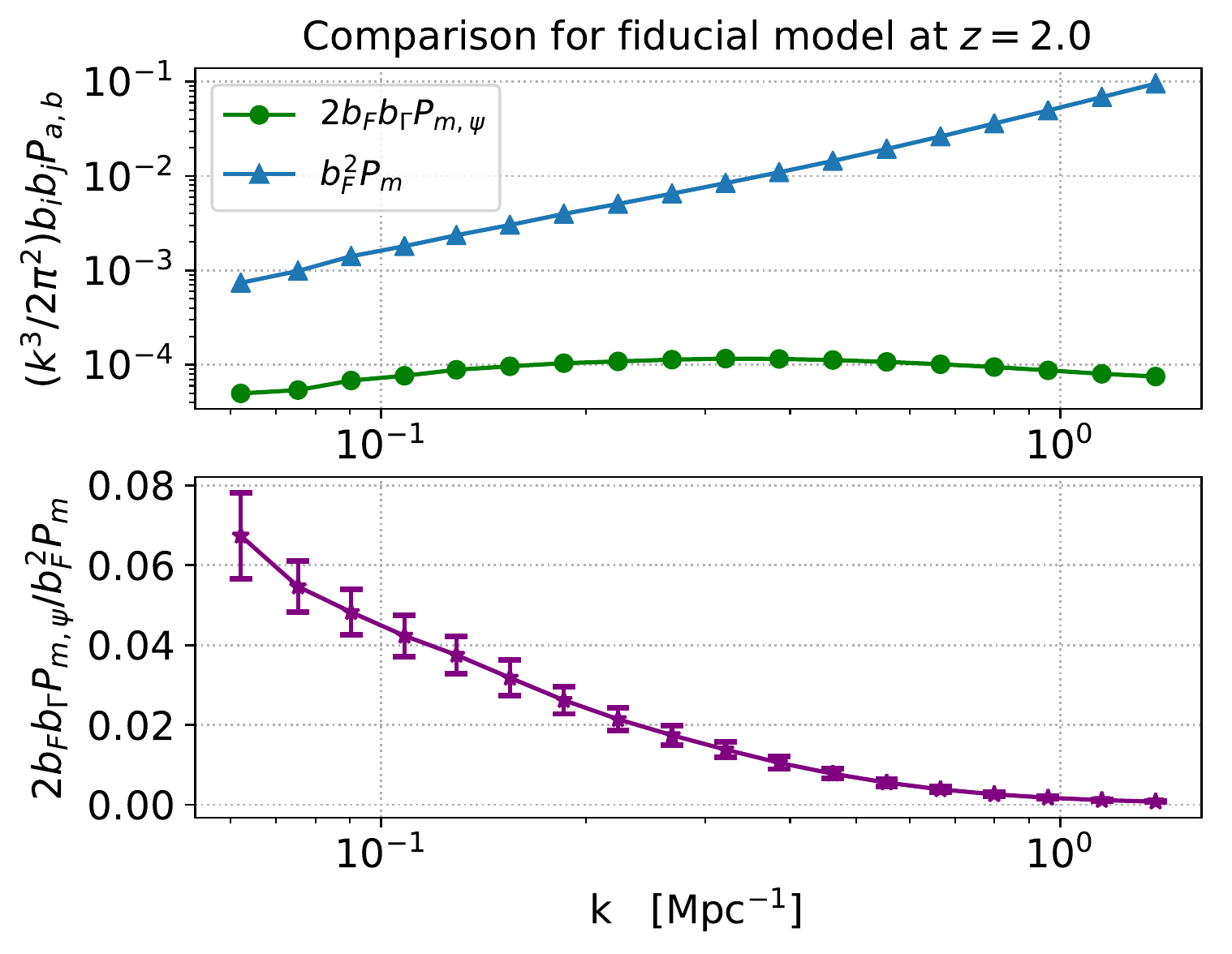}
    \caption{Impact of patchy reionization in the 3D \lya forest power spectrum at lower redshift $z=2.0$ for our fiducial model. (Top) the cross-power spectrum of matter and $\psi$ (green circles) and the auto-power spectrum of matter (blue triangles), multiplied by their respective relevant prefactors, as a function of wavenumber $k$. The matter power spectrum is computed using {\sc CLASS} \citep{2011JCAP...07..034B}. The error bar in $P_{m,\psi}$ is the sample variance in our simulations and is smaller than the chosen scale of circles. (Bottom) the deviation of the 3D \lya power spectrum due to patchy reionization, i.e.\  the ratio of the second term over the first term of Eq.~(\ref{eq:p3D}) which equals to $2  (b_{\Gamma}/b_{\rm F}) P_{\rm m, \psi}/P_{\rm m} $. Note that $P_{m,\psi}$ is negative, $b_{\rm F}$ is negative, and $b_\Gamma$ is positive, so the deviation is positive.}
    \label{fig:mp1}
\end{figure}

\begin{figure}
    \centering
    \includegraphics[width=\columnwidth,keepaspectratio]{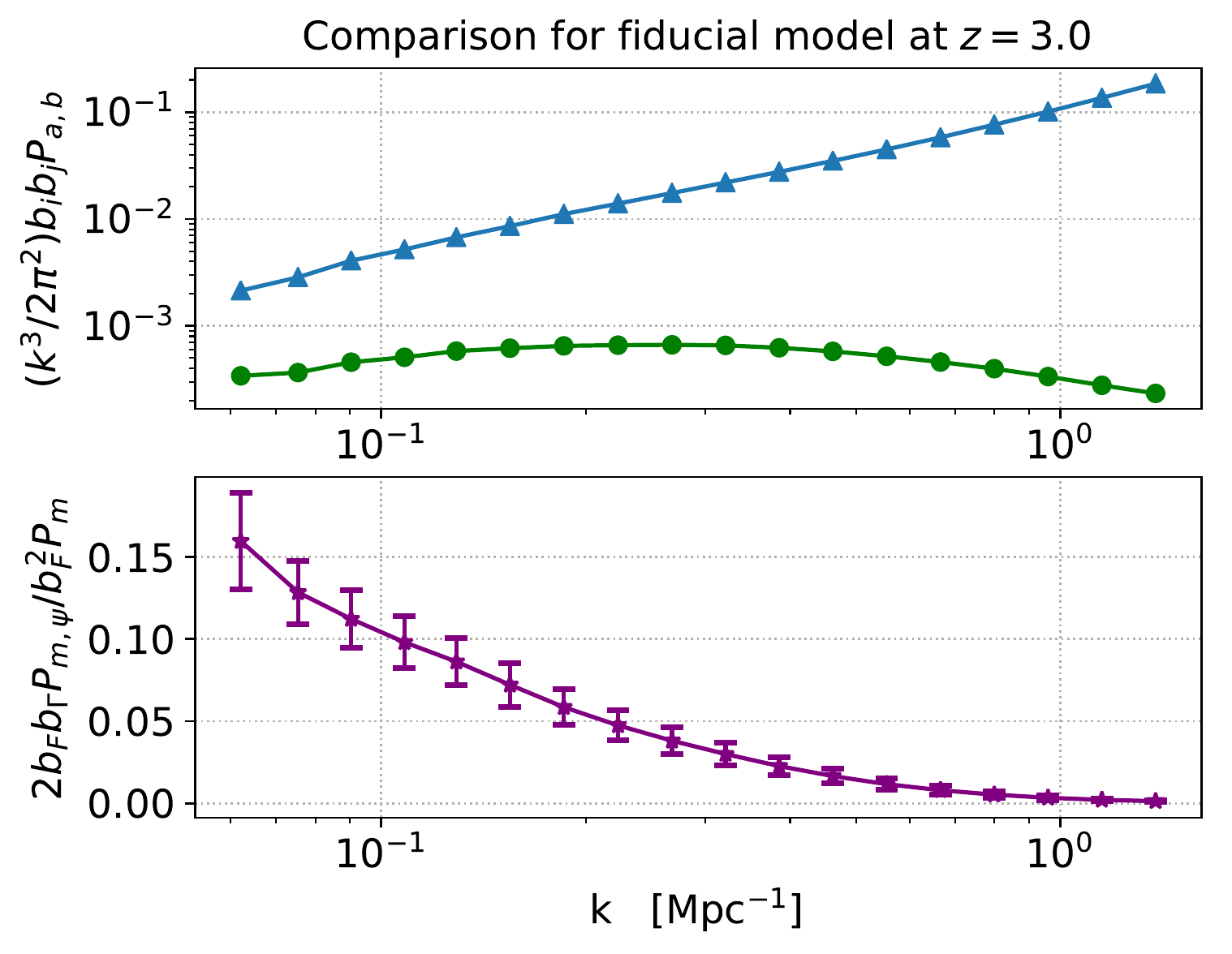}
    \caption{Same as Figure~\ref{fig:mp1} but at $z=3.0$.}
    \label{fig:mp3}
\end{figure}

\begin{figure}
    \centering
    \includegraphics[width=\columnwidth,keepaspectratio]{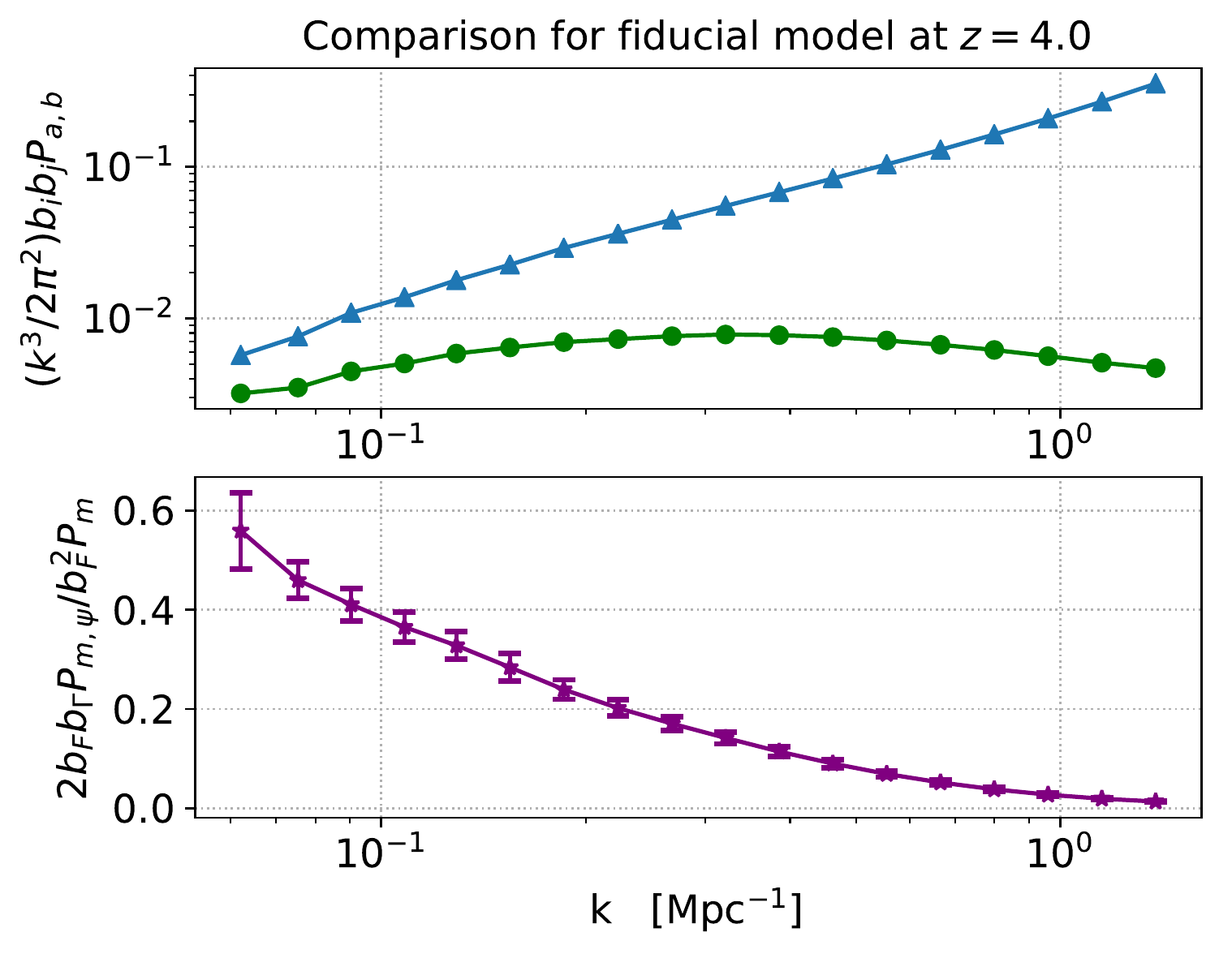}
    \caption{Same as Figure~\ref{fig:mp1} but at higher redshift $z=4.0$.}
    \label{fig:mp5}
\end{figure}

In this section we illustrate how strong the impact of patchy reionization in the \lya forest is. For clarity, we first show the integrand in Eq.~(\ref{eq:psicross}) as a function of redshift, for our fiducial model. Figure \ref{fig:fid_int} shows that the memory of inhomogeneous reionization in the \lya forest is mainly sensitive to the midpoint and later stages of reionization ($\bar{x}_{\rm HI} \approx 0.25$). However, the thermal relics are not particularly sensitive to the final stage of reionization (e.g. $\bar{x}_{\rm HI} = 0.10$ corresponding to $z = 6.29$ for our fiducial model). This trend might be confusing given that naturally one would expect the memory from reionization to be stronger at the end of reionization. However, as \emph{inhomogeneous} reionization completes, the Universe is almost completely filled by overlapping bubbles. Hence the perturbations in $x_{\rm HI}$ go to zero. Furthermore, the bumpiness in our integrand is a consequence of using only 7 sudden reionization scenarios ($z_{\rm re} = \{6,7,8,9,10,11,12\}$) in our small-scale simulations. An implementation of Eq.~(\ref{eq:psicross}) for precision cosmology will require more computational resources allocated to run more sudden reionization scenarios in small-scale high-resolution simulations capable of tracking the small-scale structure.

In Table \ref{tab:result}, we report the percentage deviation of the 3D \lya power spectra due to patchy reionization, i.e.\ the ratio of the second term over the first term of Eq.~(\ref{eq:p3D}) multiplied by $100\%$ which equals to $2  (b_{\Gamma}/b_{\rm F}) P_{\rm m, \psi}/P_{\rm m} \times 100\% $, at $k = 0.14 \ \textup{Mpc}^{-1}$ at various redshifts for the different models included in this study. Furthermore, the 1D \lya power spectrum has already been measured by observations, and hence we use the latest data release by BOSS+eBOSS \citep{2019JCAP...07..017C} to estimate the degree of contamination. Hence, for the 1D power spectrum, we tabulate $2b_{\rm F}b_{\rm \Gamma}P_{m,\psi}^{\rm 1D}/P_{\rm F}^{\rm eBOSS} \times 100\%$. 

We observe an increase of the significance of the effect of patchy reionization in the 1D \lya forest power spectra (see Figure \ref{fig:fid_1D}) overall compared to the previous results from \cite{2019MNRAS.487.1047M}. The primary reason for this is that we corrected a typo in the integration method from 3D to 1D power spectrum in the previous work of \cite{2019MNRAS.487.1047M}. Other factors that contribute to a smaller degree is the use of the latest, improved, eBOSS measurements and the astrophysics of reionization explored in this work. Our results as seen in Figure \ref{fig:fid_1D} agree with Figure 7 of \cite{2019MNRAS.486.4075O}. We note that even if our theoretical 1D power spectrum reproduces the observed data reasonably well, it is still incomplete in the sense that it only includes the linear Kaiser effect and assumes that the RSD parameter has no redshift evolution and it is equal to 1 (also see \S \ref{ssec:limi}). Thus, we choose to use the observed power spectrum for the tabulated main results of this work. In particular, for $k = 0.14 \ \textup{Mpc}^{-1}$, at low redshift ($z_{\rm obs} = 2.0$) the change in the 1D \lya forest spectrum due to patchy reionization is about a per cent (ranging from $\sim 0.78 \%$ in our T3 model to $\sim 1.40 \%$ in the R2 model). In contrast, at the high redshift ($z_{\rm obs} = 4.0$), the change is about tens of per cent (ranging from $\sim 11.0\%$ in the T3 model up to $\sim 18.9\% $ in the R2 model).

Although the impact of inhomogeneous reionization in the \lya forest is slightly larger than in our previous work, if one takes into account the statistical error in the eBOSS measurement (plotted in Figure \ref{fig:fid_1D}) and the systematic error in modelling reionization, it is currently challenging to use the 1D \lya power spectrum to extract information from the reionization epoch via the impact of inhomogeneous reionization in the \lya forest. However, in future work we will investigate this possibility.


On the other hand, the expected effect of patchy reionization in the 3D \lya forest power spectrum are significantly larger. We find that the percentage deviation is in the order of tens of per cent at $z_{\rm obs}=4.0$ (e.g.\ $\sim 51.3\%$ for our model R2), and in the order of a few per cent at $z_{\rm obs}=2.0$ (e.g.\ $\sim 3.19\%$  for our model T3). 

Our results for the survival of these thermal relics from reionization to low redshift might seem \emph{unconventional}. However, the reason we obtain such strong ripples in the \lya forest power spectra is because we are able to carefully track the small-scale structure prior to reionization. In particular, mini-voids at reionization get later compressed to mean-density while they simultaneously get reionized to high entropy. This High-Entropy Mean-Density (HEMD) gas gets usually heated to $\sim 3 \times 10^4 \ \textup{K}$ in our simulations and remains above the temperature-density relation even at $z<4$. As a result the temperature-density relation of the IGM becomes bimodal with the usual low-entropy phase and this high-entropy phase. The ultimate fate of the HEMD phase is to merge with the low-entropy mode; however, it takes cosmological time scales for the merging to complete (see Figure 4 and 5 of \citealt{2018MNRAS.474.2173H}).

We stress here that it is really both the patchy nature of reionization $P_{m,x_{\rm HI}}$, and the HEMD gas, i.e. $\partial \psi /\partial z$ in Eq.~(\ref{eq:psicross}) \emph{together} that result in the memory of reionization in the \lya forest computed in this work. To illustrate the former point, we turn off the inhomogeneous recombination prescription of \cite{2014MNRAS.440.1662S} in {\tt 21cmFAST}, and consider the resulting, faster, reionization. Our results for the 3D \lya power spectrum are reduced by a factor of $2$ at $z = 2.0$ and $\sim 1.8$ at $z = 4.0$. This decrease of deviations in the \lya power spectrum is expected since a faster reionization model will have a narrower range of redshifts for the peak of $P_{m,x_{\rm HI}}$ that contribute the most to the integration in Eq.~(\ref{eq:psicross}). Regarding the importance of the HEMD gas, on the other hand, we artificially limit the impact of the HEMD gas by only considering our transparency results for higher redshifts, say $z_{\rm re} = \{8.0, 9.0, 10.0, 11.0, 12.0 \}$. Note that patches of the sky that reionize earlier will have the high entropy gas merge with the low entropy gas sooner, thus recovering the temperature-density relation of the IGM faster. In this case we obtain a decrease in the 3D \lya results with a factor of $2.2$ at $z = 2.0$ and $1.8$ at $z = 4.0$. 

In Figure~\ref{fig:mp1}, we plot the $k$-dependence of both the auto-power spectrum of matter and the cross-power spectrum of matter and the change of transparency of the IGM due to inhomogeneous reionization, at redshift $2.0$. In the lower panel, we show the ratio of the contribution to the flux power spectrum from inhomogeneous reionization over the cosmological contribution. Similarly, we illustrate the evolution at $z_{\rm obs} = 3.0$ in Figure~\ref{fig:mp3}, and at $z_{\rm obs} = 4.0$ in Figure~\ref{fig:mp5}.

Even though the impact of patchy reionization in the \lya forest is significant --- especially at higher redshift since the IGM have not had enough time to relax to the usual temperature-density relation --- on the large scales, the effect diminishes for the small scales. This behaviour is expected because the memory of patchy reionization couples to the reionization bubble scales. The different $k$-dependence of the matter power spectrum and that of $P_{m,\psi}$ is a positive sign for modeling and extracting, or marginalizing over, this broadband signal. 

For reference, the statistical error per bin of the eBOSS $P^{\rm 1D}_{\rm F}$ measurement for $k = 0.151 \ \textup{Mpc}^{-1}$ are 1.05 per cent at $z = 2.6$, 1.23 per cent at $z = 3.0$ and 6.03 per cent at $z = 4.0$. Even in the hypothetical scenario where the statistical error budget of DESI for 1D power spectrum measurements would be only a half of the recent BOSS+eBOSS \citep{2019JCAP...07..017C} measurements, the memory from inhomogeneous reionization would be significantly larger than the statistical error at $z=4$. In fact, DESI is very likely to manage a much better measurement. In that more promising case, this signal would be significant for most of the models in Table \ref{tab:result}. 

As seen in Eq.~(\ref{eq:psicross}) this broadband systematic effect for the \lya forest is fundamentally linked to the astrophysics of reionization, and hence indirectly coupled to the physics of the cosmic dawn. Given the current capabilities of instruments like DESI \citep{2016arXiv161100036D} and 4MOST \citep{2019Msngr.175...50R}, the 3D \lya forest will be measured in a time span of a couple of years. In the absence of any mitigation scheme, theoretically, one should be able to use this effect to construct a new avenue for constraining the reionization and thermal histories once DESI has measured the 3D \lya power spectrum. We will explore this plausible scenario in future work.

\begin{figure}
    \centering
    \begin{minipage}{\linewidth}
\centering
\includegraphics[width=\columnwidth]{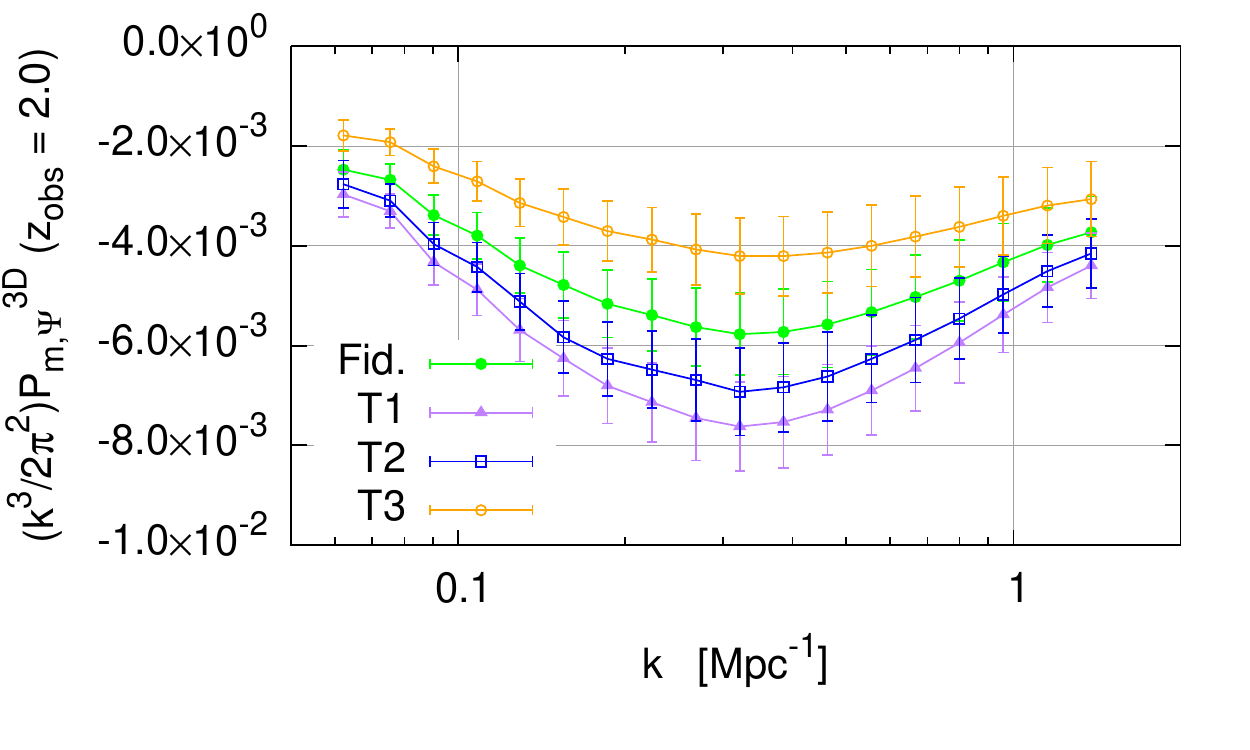}
\end{minipage}
\begin{minipage}{\linewidth}
\centering
\includegraphics[width=\columnwidth]{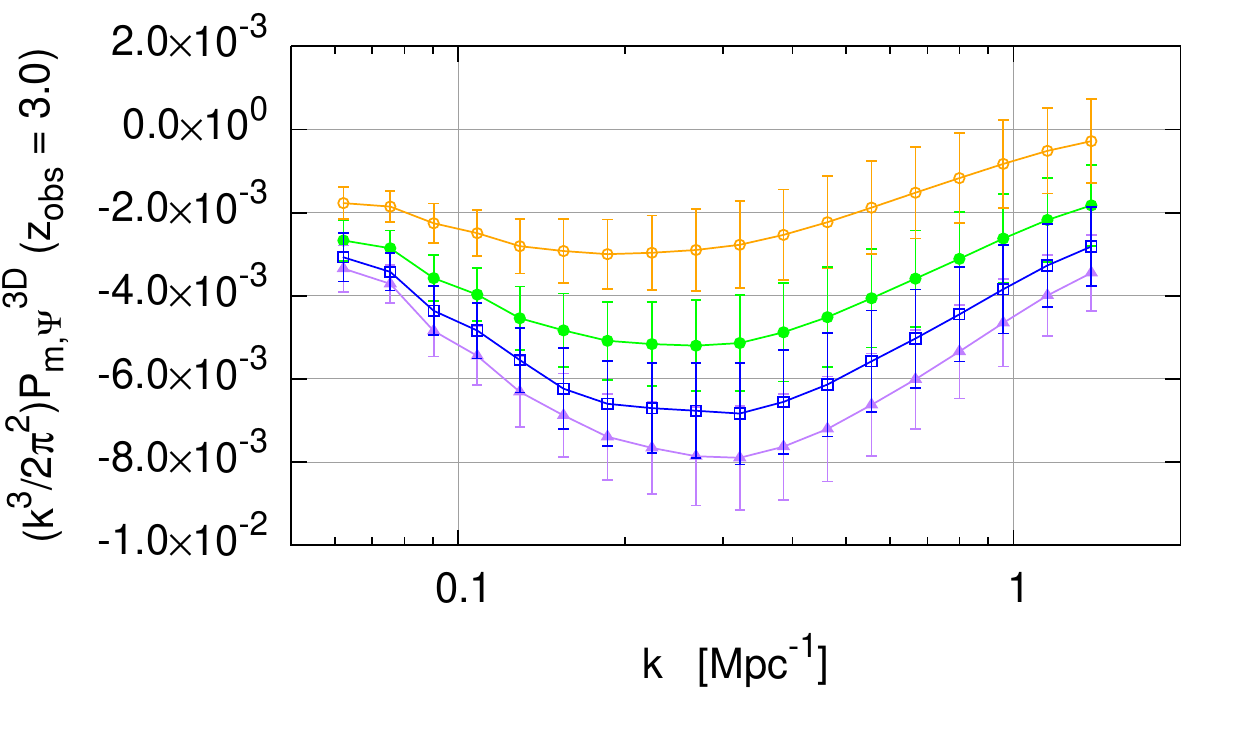}
\end{minipage}
\begin{minipage}{\linewidth}
\centering
\includegraphics[width=\columnwidth]{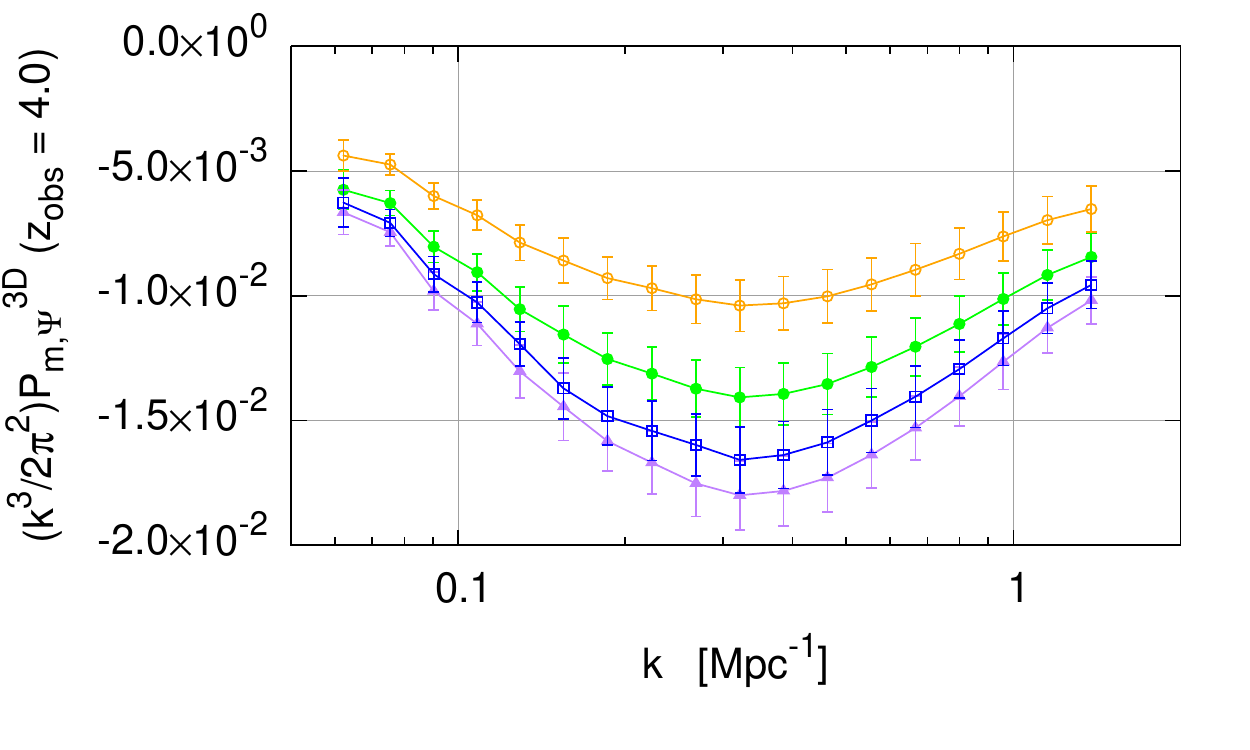}
\end{minipage}
    \caption{Cross-power spectrum of matter density and the transparency of the IGM, $P_{m,\psi}$, for the \lya forest observed at redshift $z_{\rm obs} = 2.0$ (top), 3.0 (middle), and 4.0 (bottom), respectively. In each panel, we consider the ``T'' models wherein the minimum virial temperature of haloes that host ionizing sources takes the value of $T_{\rm min} = 5 \times 10^4\,{\rm K}$ (T1 model, purple), $4 \times 10^4\,{\rm K}$ (T2 model, blue), and $2 \times 10^4\,{\rm K}$ (T3 model, orange), in comparison with our fiducial model wherein $T_{\rm min} = 3 \times 10^4\,{\rm K}$ (green).}
    \label{fig:T_psi}
\end{figure}

\begin{figure}
    \centering
    \begin{minipage}{\linewidth}
\centering
\includegraphics[width=\columnwidth]{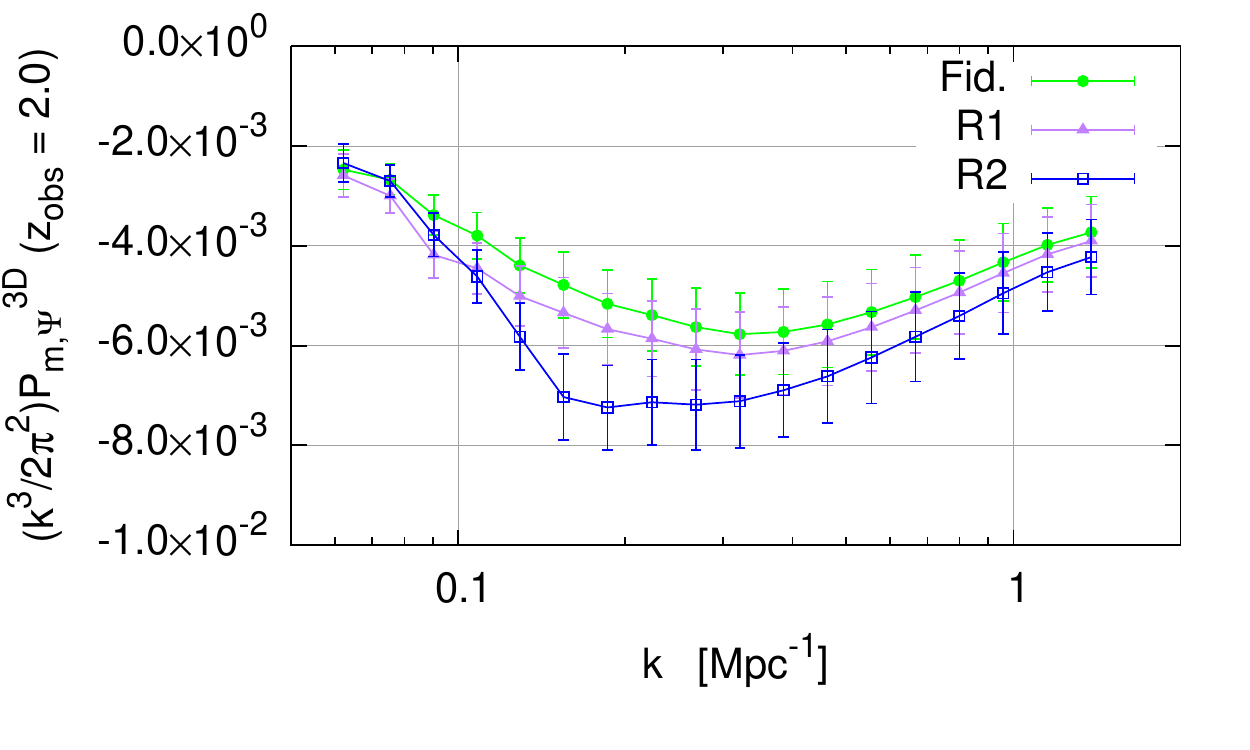}
\end{minipage}
\begin{minipage}{\linewidth}
\centering
\includegraphics[width=\columnwidth]{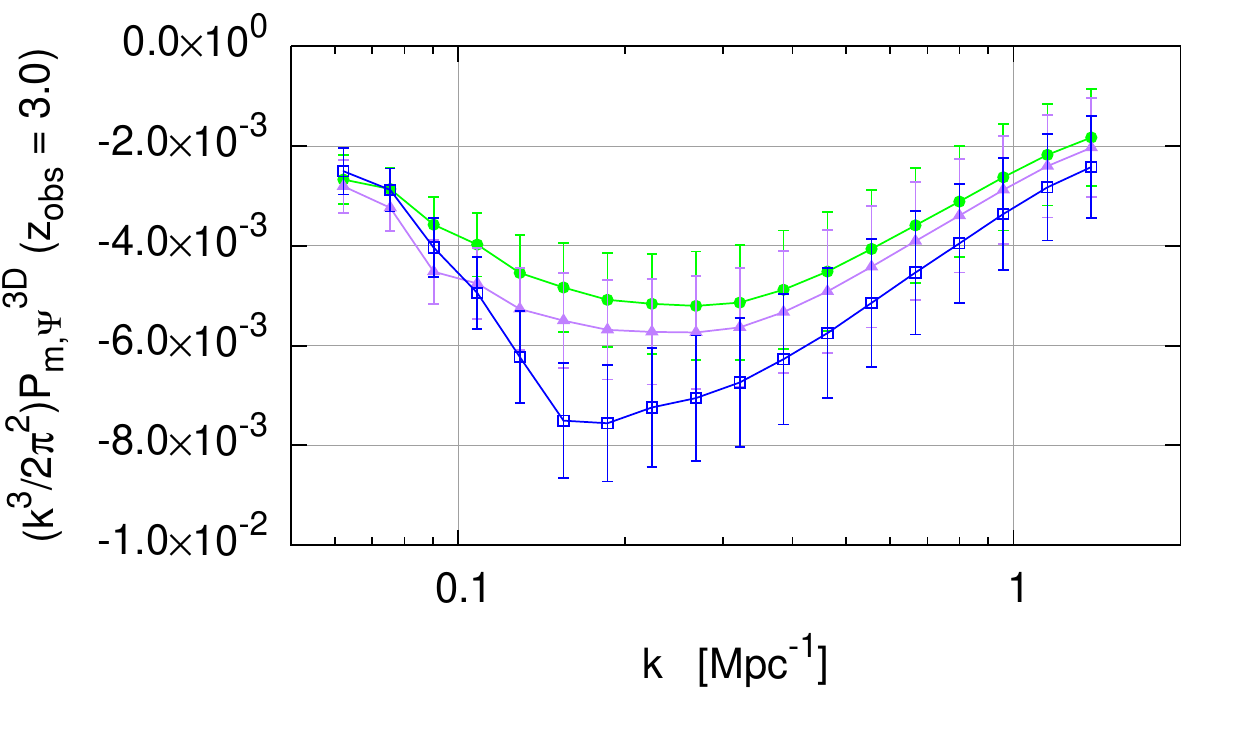}
\end{minipage}
\begin{minipage}{\linewidth}
\centering
\includegraphics[width=\columnwidth]{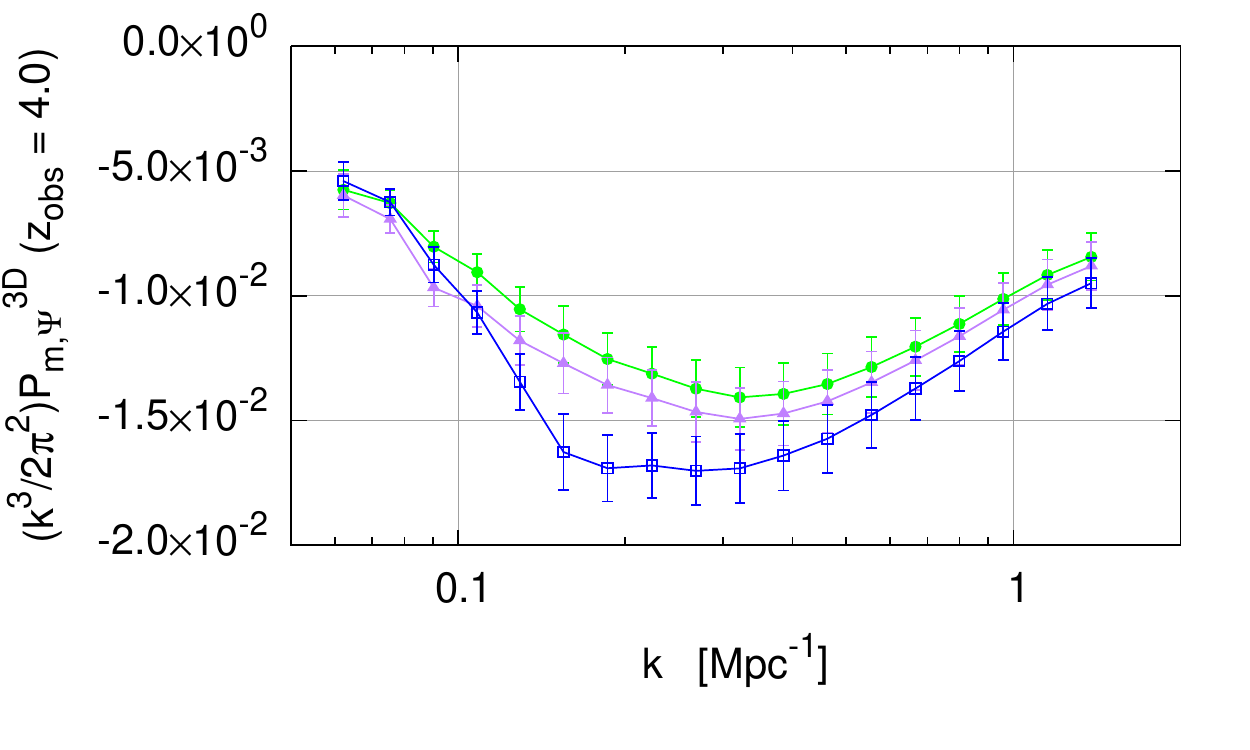}
\end{minipage}
    \caption{Same as Figure~\ref{fig:T_psi} but for the ``R'' models wherein the mean free path of ionizing photons takes the value of $R_{\rm mfp} = 25 \ \textup{Mpc}$ (R1 model, purple) and $15 \ \textup{Mpc}$ (R2 model, blue), in comparison with our fiducial model wherein $R_{\rm mfp} = 50 \ \textup{Mpc}$ (green).}
    \label{fig:R_psi}
\end{figure}

\begin{figure}
    \centering
    \begin{minipage}{\linewidth}
\centering
\includegraphics[width=\columnwidth]{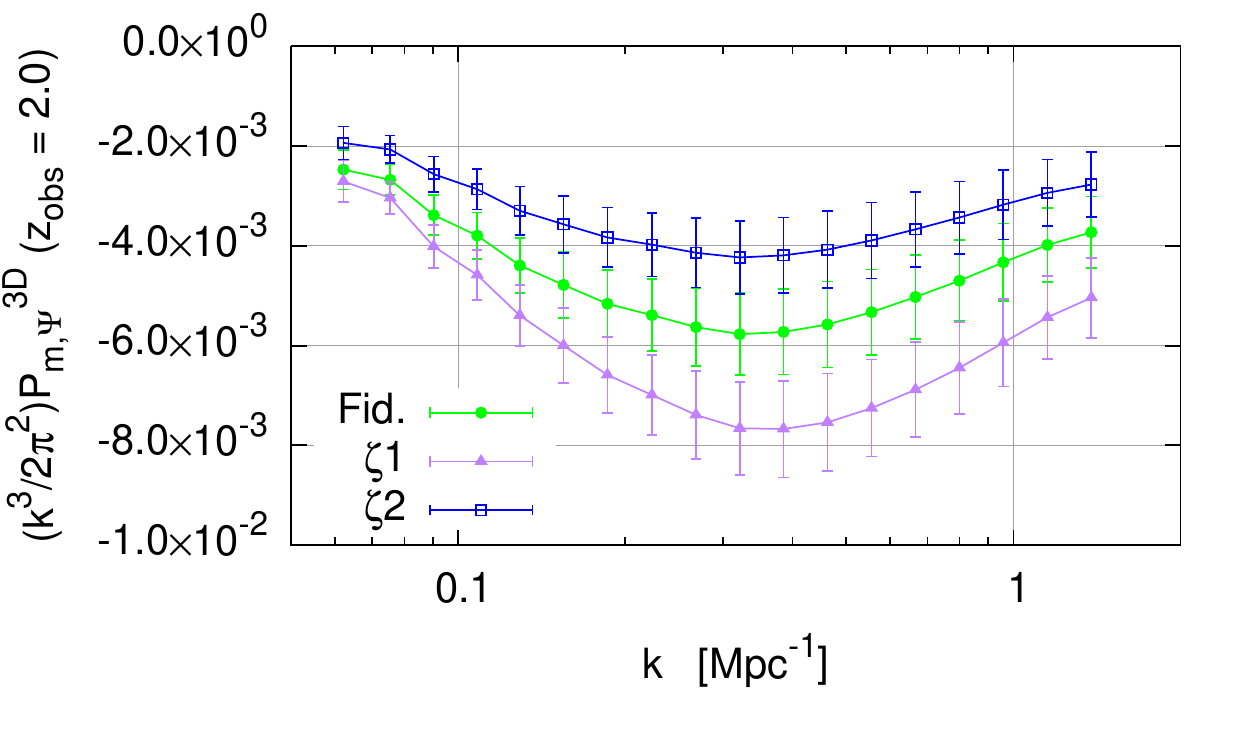}
\end{minipage}
\begin{minipage}{\linewidth}
\centering
\includegraphics[width=\columnwidth]{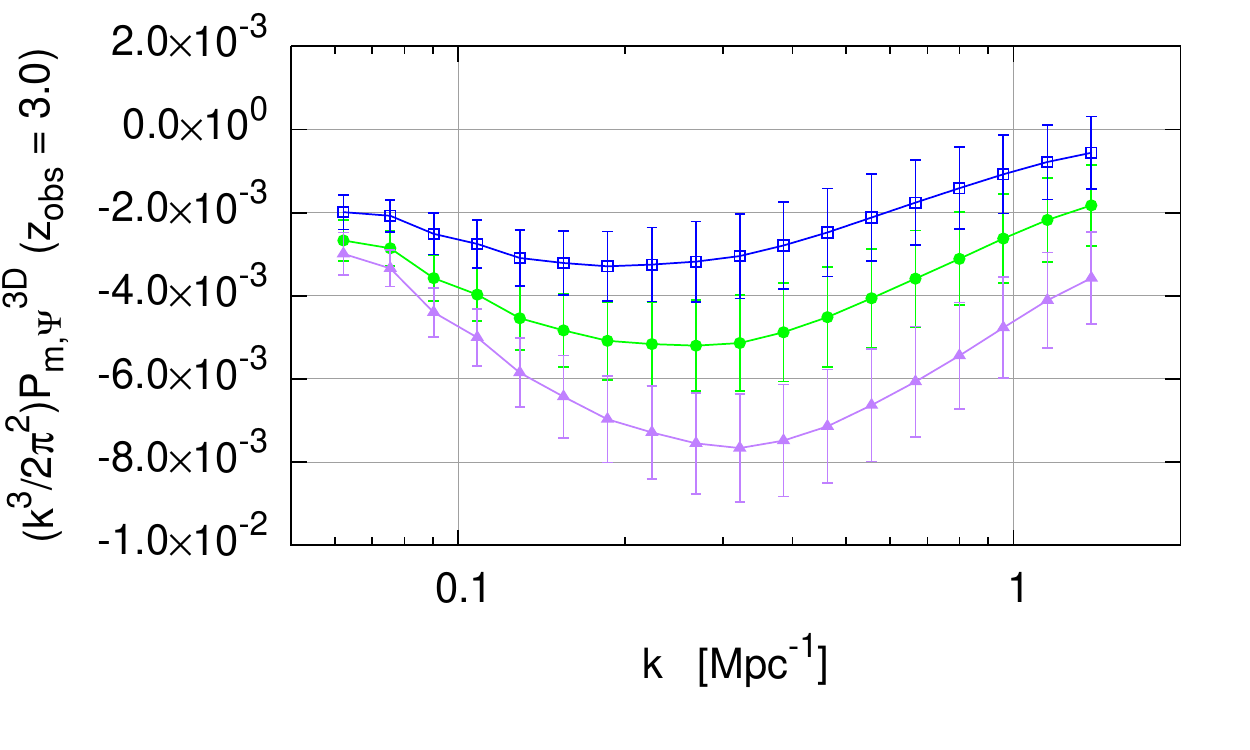}
\end{minipage}
\begin{minipage}{\linewidth}
\centering
\includegraphics[width=\columnwidth]{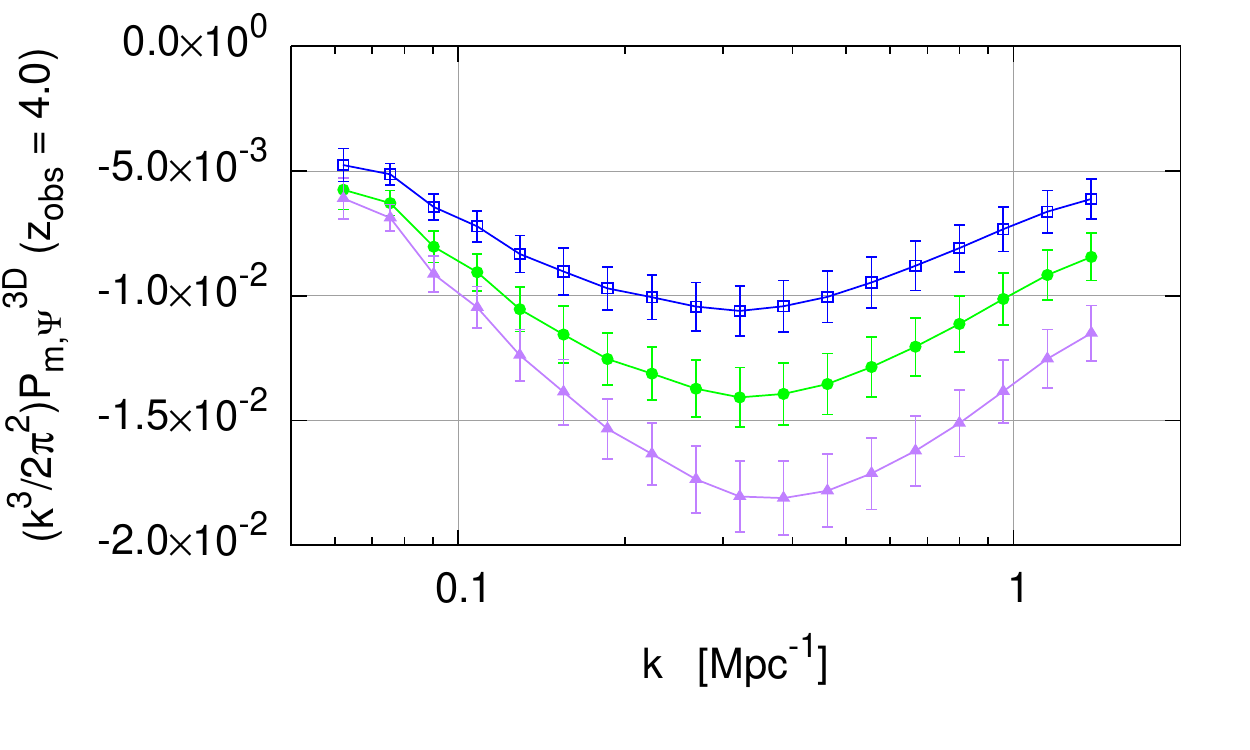}
\end{minipage}
    \caption{Same as Figure~\ref{fig:T_psi} but for the ``$\zeta$'' models wherein the ionizing efficiency takes the value of $\zeta = 20$ ($\zeta_1$ model, purple), and $30$ ($\zeta_2$ model, blue), in comparison with our fiducial model wherein $\zeta = 25$ (green).}
    \label{fig:zeta_psi}
\end{figure}

\begin{figure}
    \centering
    \begin{minipage}{\linewidth}
\centering
\includegraphics[width=\columnwidth]{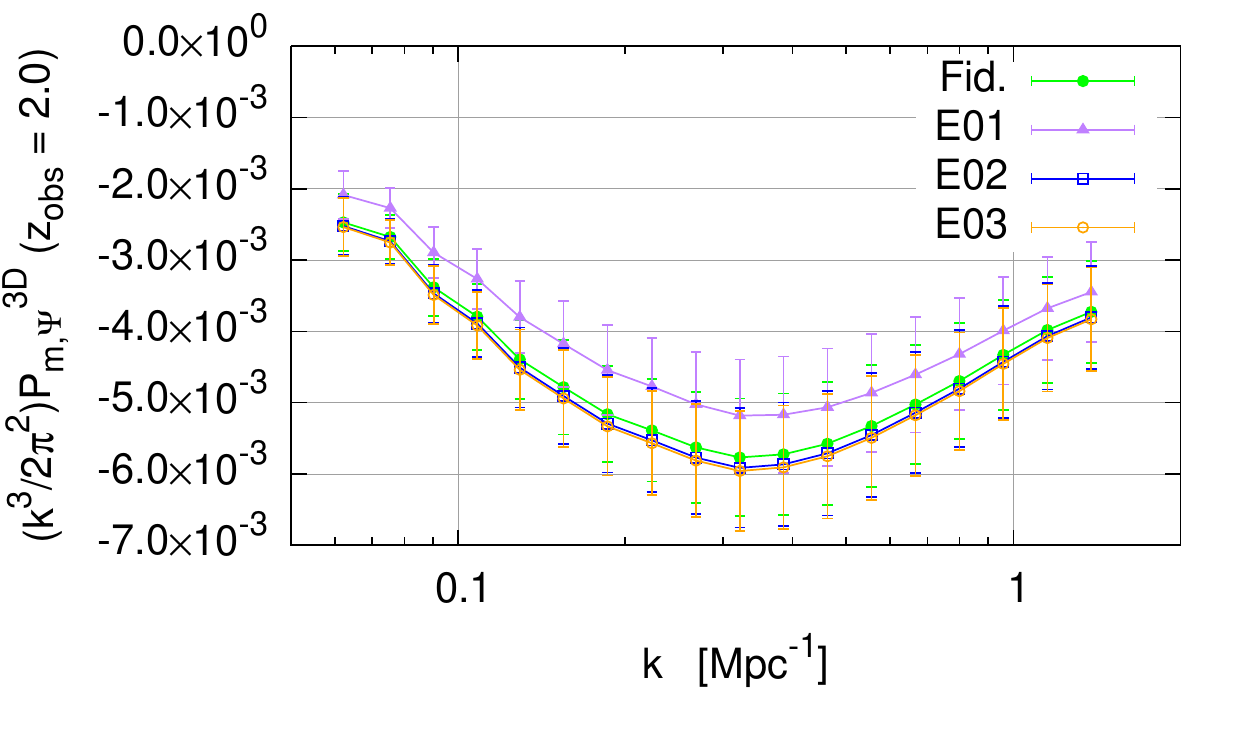}
\end{minipage}
\begin{minipage}{\linewidth}
\centering
\includegraphics[width=\columnwidth]{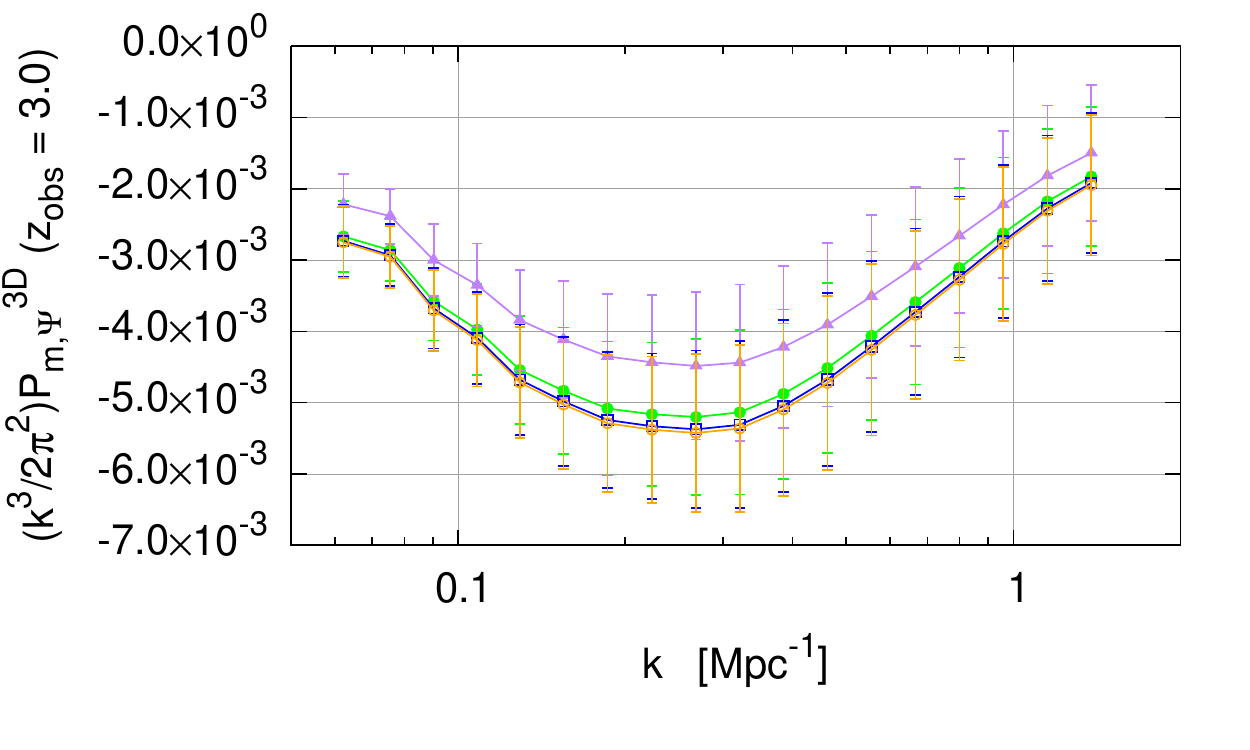}
\end{minipage}
\begin{minipage}{\linewidth}
\centering
\includegraphics[width=\columnwidth]{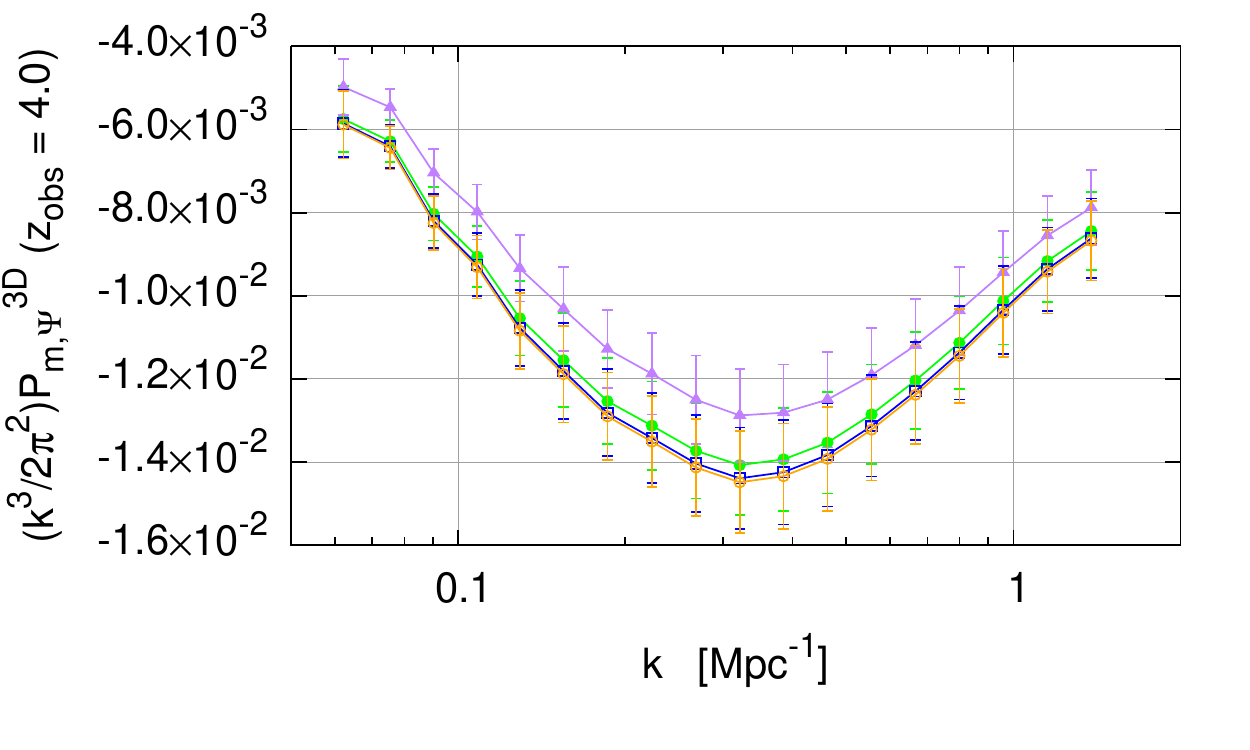}
\end{minipage}
    \caption{Same as Figure~\ref{fig:T_psi} but for the ``E0'' models wherein the energy threshold for the lowest energy X-ray photons not absorbed by galaxies takes the value of $E_0 = 100\,{\rm eV}$ (E01 model, purple), $1000\,{\rm eV}$ (E02 model, blue), and $1500\,{\rm eV}$ (E03 model, orange), in comparison with our fiducial model wherein $E_0 = 500\,{\rm eV}$ (green).}
    \label{fig:E0_psi}
\end{figure}

\begin{figure}
    \centering
    \begin{minipage}{\linewidth}
\centering
\includegraphics[width=\columnwidth]{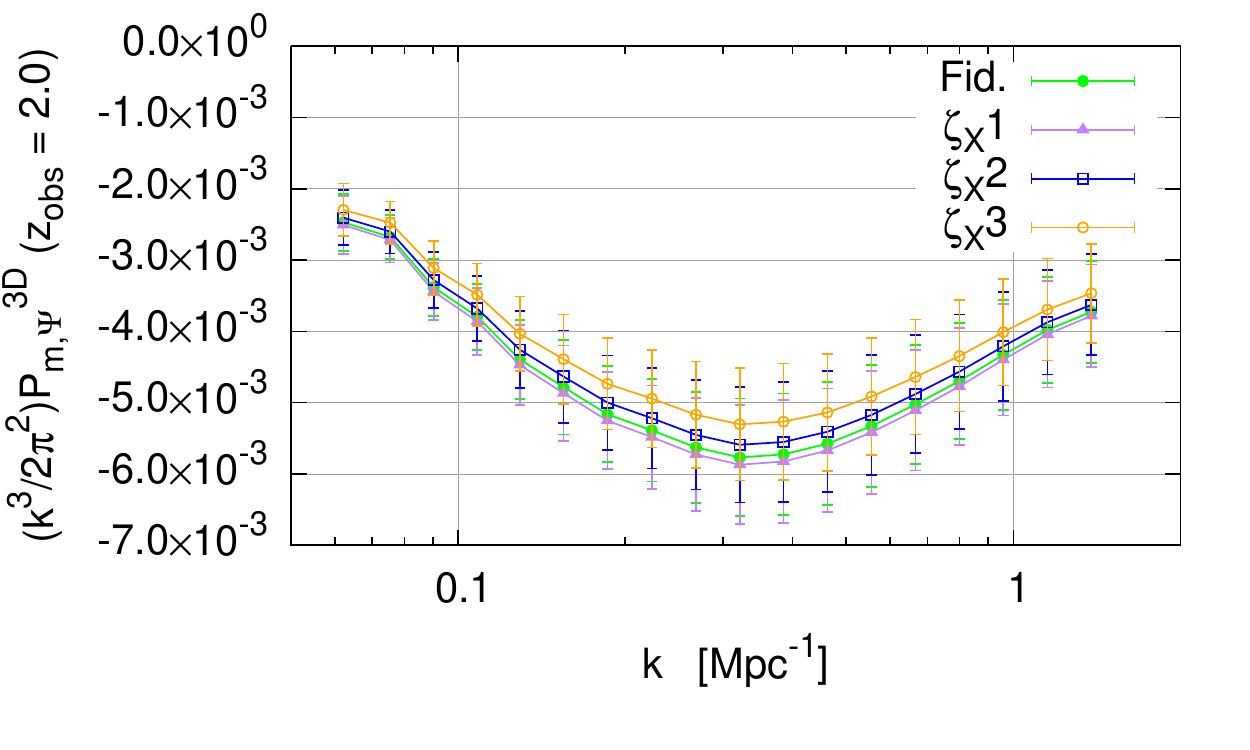}
\end{minipage}
\begin{minipage}{\linewidth}
\centering
\includegraphics[width=\columnwidth]{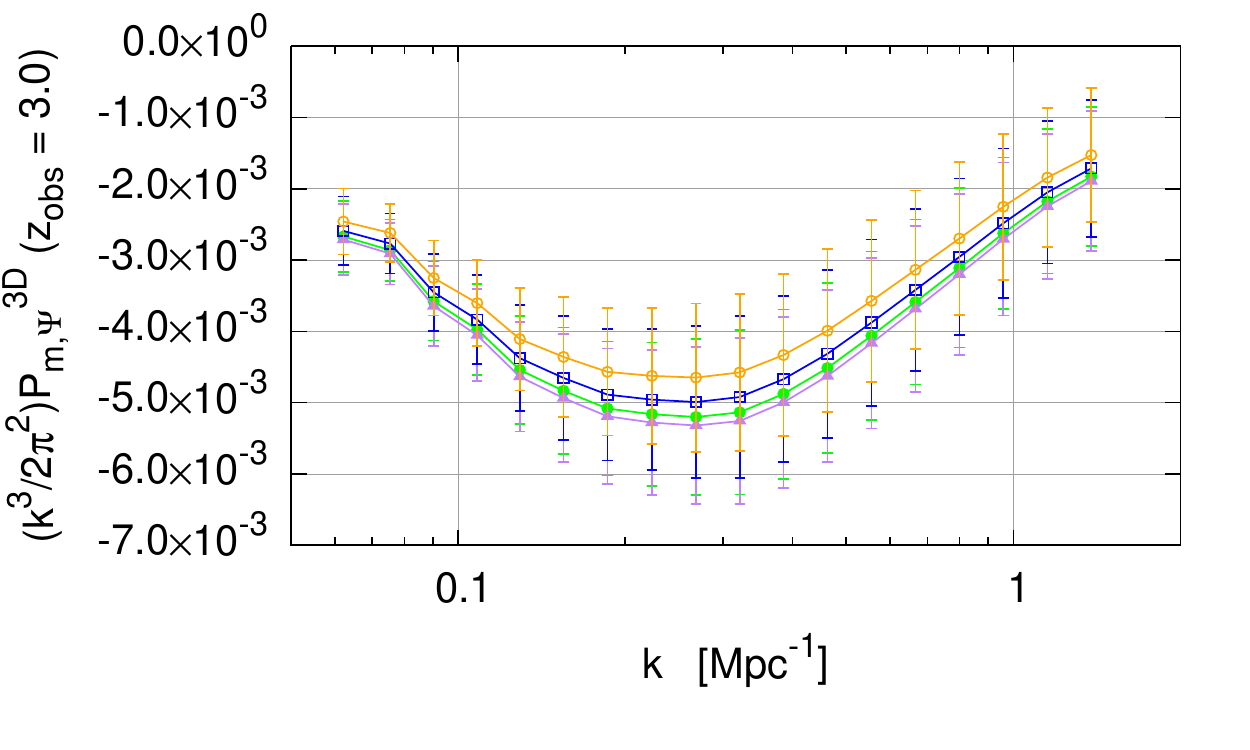}
\end{minipage}
\begin{minipage}{\linewidth}
\centering
\includegraphics[width=\columnwidth]{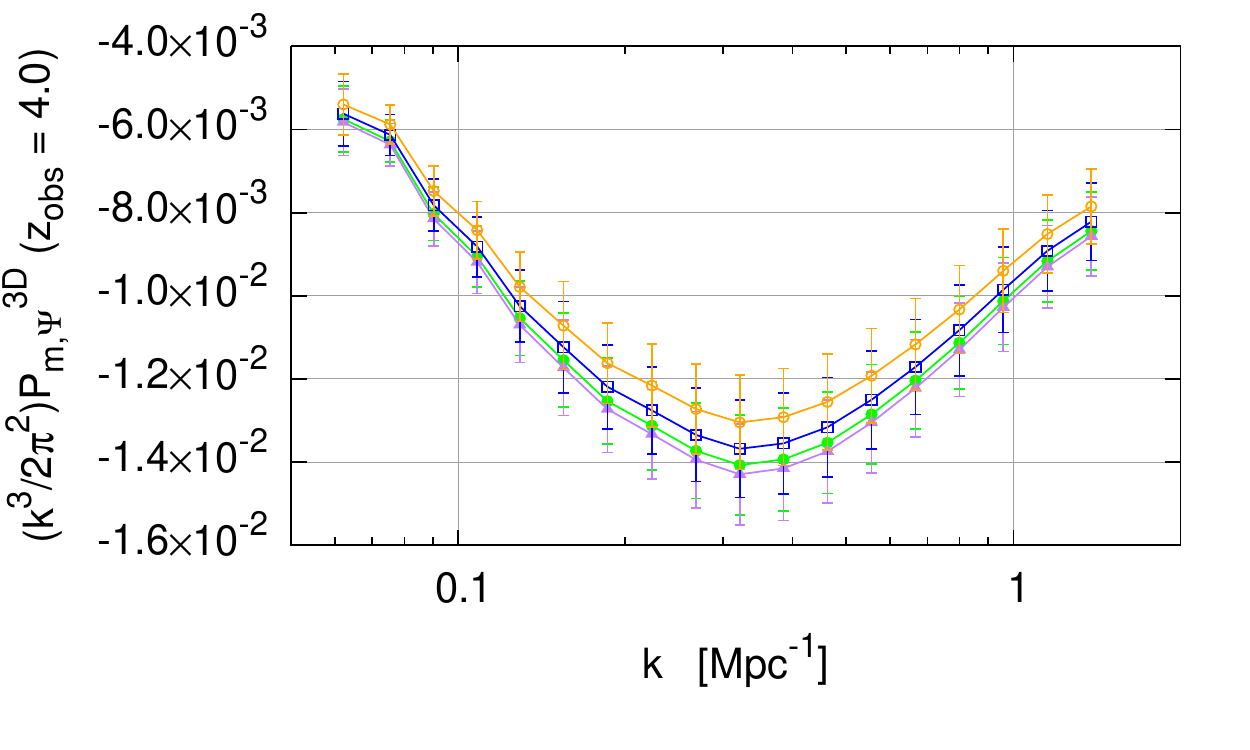}
\end{minipage}
    \caption{Same as Figure~\ref{fig:T_psi} but for the ``$\zeta_X$'' models wherein the X-ray efficiency takes the value of $\zeta_{X} = 1 \times 10^{56}\,\textup{M}_\odot^{-1}$ ($\zeta_X1$ model, purple), $4 \times 10^{56}\,\textup{M}_\odot^{-1}$ ($\zeta_X2$ model, blue), and $8 \times 10^{56}\,\textup{M}_\odot^{-1}$ ($\zeta_X3$ model, orange), in comparison with our fiducial model wherein $\zeta_{X} = 2 \times 10^{56}\,\textup{M}_\odot^{-1}$ (green).}
    \label{fig:Lx_psi}
\end{figure}

\subsection{Dependence on astrophysical scenarios}
Having shown the effect of patchy reionization in the \lya forest for our fiducial model and illustrated its significance as a systematic, we now proceed to focus on its potential as a link to the astrophysics of cosmic reionization and cosmic dawn by analyzing the dependence on the different astrophysical parameters, and also strategize how to separate cosmology from astrophysics. 

In Figure \ref{fig:T_psi}, we show the cross-power spectrum of the matter and transmission of the IGM,  $P_{m,\psi}$, as a function of wavenumber for the models that vary the minimum virial temperature of haloes with efficient star formation. Also in these panels we show the dependence of these signals on the redshift of observation. We highlight that 
it is theoretically possible to distinguish between the reionization models because the power spectra in these models show different shapes in wavenumber and different overall amplitudes. The dip of the power spectrum corresponds to the minimum of $P_{m,x_{\rm HI}}$ although smoothed by the integration and multiplied by the factors that depend on redshift of observation. We find that the dip of the cross-power spectrum in the T1 model which has the largest $T_{\rm min}$ has the largest absolute value. Basically, reionization is dominated by more massive haloes in the T1 model because of its higher temperature threshold. This results in the larger fluctuations in neutral fraction field, and therefore larger $|P_{m,x_{\rm HI}}|$. 

Besides, increasing the value of $T_{\rm min}$ results in a later midpoint of reionization and in a faster reionization process, e.g. our T1 model has $\Delta^{T1}_{\rm re} = 2.10$, where we define the duration of reionization $\Delta_{\rm re} \equiv z(\bar{x}_{\rm HI} = 0.25) - z(\bar{x}_{\rm HI} = 0.75)$,  in comparison to $\Delta^{T3}_{\rm re} = 2.48$. In the language of Figure \ref{fig:fid_int}, the midpoint and 75 per cent completion (and later stages) of reionization align better with the maximum of the integrand for the T1 model than the other T models. Hence, the change in the memory of patchy reionization in the \lya forest is primarily due to the difference in the ionization fluctuations around reionization. We remind the reader that we are not changing our small-scale simulations and as a result all models experience the same temperature-density evolution when computing $\psi$. 

Furthermore, we show the $P_{m, \psi}$ power spectrum for the models that change the maximum allowed size of the bubbles in Figure \ref{fig:R_psi}. Generally, the dip of the cross-power spectrum becomes broader for smaller radii. Taking into account the error bars, these models exhibit similar strength in the large-scale, which is due to the similarity of the $P_{m,x_{\rm HI}}$ for the R models\footnote{We remind the reader that the $R_{\rm mfp}$ is changed here only in the {\tt 21cmFAST} simulations, i.e. for EOR redshifts -- when ionization fluctuations are not homogeneous -- and it impacts the forest through the thermal relics from reionization, particularly through the HEMD phase of the temperature-density evolution.} at these scales. However, as we move to smaller scales ($k \gtrsim 0.1 \ \textup{Mpc}^{-1}$) we see a larger deviation seeded by the difference of how reionization finishes for these three models, i.e. by the process of how many bubbles are needed to evolve and completely percolate the Universe. This deviation is present in the cross-power spectrum of matter and neutral hydrogen fraction, and it is quite small around the mid-point of reionization of these models but becomes slightly stronger as we move to lower redshifts. Moreover, the integration in Eq.~(\ref{eq:psicross}) enhances the dip resulting in the feature present in the R2 model. While the fiducial model has double the maximum allowed bubble size with respect to the R1 model, the deviation between these cross-power spectra is significantly smaller than that between the R1 and R2 models. This is due to similar reionization histories since changing $R_{\rm mfp}$ does not alter the neutral hydrogen fraction strongly. In particular the initial first half of reionization is very similar because the bubbles are too small to impact the filling of the boxes. Therefore, the deviations are due to changes in the size of the \ion{H}{II} regions themselves mainly in the second half of reionization. In particular, we have $z(\bar{x}_{\rm HI} = 0.25) = 6.79$ for the fiducial model and $z(\bar{x}_{\rm HI} = 0.25) = 6.73$ for R1, while the R2 model has a larger duration due to struggling to percolate the universe with smaller bubbles, $z(\bar{x}_{\rm HI} = 0.25) = 6.53$. However, this is not the only special feature of R2. 

Given that $R_{\rm mfp}$ is utilized in the filter radius for the ionized bubbles in the {\tt 21cmFAST} code, we note that the cross-power of matter and bubble spatial structure in the R2 model might have artificial power given by increased shot-noise due to some isolated regions included into a filter. Therefore, the real behavior of $P_{m,\psi}$ with maximum bubble size might resemble more the changes between the R1 and the fiducial model.


The cross-power spectra $P_{m, \psi}$ for the models that vary the ionizing efficiency are shown in Figure \ref{fig:zeta_psi}. The most interesting feature is the difference of $P_{m, \psi}$ between the models, i.e.\ the amplitude of $P_{m, \psi}$ is very sensitive to the value of the ionizing efficiency $\zeta$. We see the expected hierarchical structure due to the $\zeta$ values, i.e. one expects the $\zeta_2$ model to have a less prominent dip than the fiducial model. Naturally, we also observe significant differences of more than 28 per cent between the fiducial model and the $\zeta_1$ model for $P_{m, \psi}$. This deviation is mainly due to the fact that the ionization efficiency dominates the reionization timing in the reionization simulation, i.e. the $\zeta_1$ model has both a delayed midpoint of reionization and a lower redshift $z = 6.26$ at $\bar{x}_{\rm HI} = 0.25$. In contrast to the T models with delayed reionization, the duration $\Delta^{\zeta_1}_{\rm re} = 2.42$ compared to $\Delta^{T1}_{\rm re} = 2.10$ is not shortened because the same number of UV photons are available to reionize the universe for the T1 and fiducial model; however, photons in the T1 model have a harder time escaping their host galaxies. For reference, the $\zeta_2$ model with earlier reionization has $z(\bar{x}_{\rm HI} = 0.25) = 7.22$ and $\Delta^{\zeta_2}_{\rm re} = 2.22$. Again, we see that parameters that influence the ionization fluctuations the most have the most impact in the \lya forest power spectrum. 

In Figure \ref{fig:E0_psi}, we illustrate the change of $P_{m,\psi}$ with varying energy threshold for X-ray photons to be not absorbed by galaxies. We see small deviations from the fiducial model, even for the extreme scenario of photon-abundance preheating of the IGM (E01 model); in particular, we see negligible changes for the photon-starved preheating of the IGM (E03 model). Although the photon-abundance scenario does have a slightly faster reionization process compared to the other models in the plot ($\Delta_{\rm re}^{\rm E01} = 2.19$ in contrast to $\Delta_{\rm re}^{\rm E03} = 2.32$), its integrand covers less of the maximum peak, which translates into a smaller impact in $P_{m,\psi}$ than that of the other models. Similarly, in Figure \ref{fig:Lx_psi} where we show the analog plots for the efficiency of X-ray photons that manage to escape their host galaxies, the $\zeta_X$ models are effectively indistinguishable in this signal, especially at lower redshifts. This is due to quite similar reionization histories both in midpoint and duration, and also because the $\zeta_X$ models do not vary the ionization fluctuations directly, their main role is to set up the initial conditions to the EOR by modifying the heating process. Therefore, we conclude that the heating models do not necessarily play a key role in the effect of patchy reionization on the \lya forest. 

We note that all models studied here generically show the similar ``smiley face'' in the cross-power spectrum $P_{m, \psi}$ as a function of wavenumber. However, the details of the dip of the power spectrum, including the shape information (such as the (a)symmetry and the width) and the amplitude, do depend on the astrophysical parameters, particularly parameters of the bubble models. The difference of $P_{m, \psi}$ is a promising sign of the potential for this systematic signal to become a constraint mechanism for the physics of cosmic reionization. In principle, one could attempt to include astrophysical parameters of reionization to the fit of the \lya forest 3D power spectrum data to also constrain the midpoint and later parts of the reionization process; however, the efficiency of this extraction remains to be seen. Quantifying this efficiency is left for future work.  

However, there are two important caveats. First, $P_{m,\psi}$ is ``robust'' against changes of the preheating of the IGM as long as one does not allow for very extreme scenarios, i.e. a really extreme X-ray photon starvation thermal history or an extreme over-abundance. 
The second caveat could be deeply influential. In this work we have analyzed only the astrophysical parameter space of the large-scale reionization simulations, and currently missed the effect of preheating in the IGM in the small-scale high resolution simulations. Nevertheless, \cite{2018MNRAS.474.2173H} did an initial estimate of the effect of X-ray preheating in the transparency of the IGM. In particular, they ran a test model with X-ray preheating for a simulation with box size of 425 kpc with $2\times(128)^3$ number of particles, and with dark matter and gas particle masses of $1.21 \times 10^3\,\textup{M}_\odot$ and $2.27 \times 10^2 \ \textup{M}_\odot$, respectively. For their late instantaneous reionization scenario ($z_{\rm re} = 7.0$), they obtained a change in the transparency of the IGM at $z_{\rm obs} = 4.0$ of 7.92 per cent with X-ray preheating in contrast to 8.20 percent with no X-rays\footnote{In comparison, we find a change of 9.28 per cent at the same redshift for our \emph{larger} small-scale simulations with no X-ray prescription.}. The presence of X-rays suppresses the effect of reionization on the IGM due to the preheating wiping out some of the small-scale structure that otherwise would be present, and hence certain patches of gas will relax faster into the usual temperature-density relation. Hence the small-scale structure is more sensitive to the effect of X-ray preheating in the IGM than the large-scale structure. 

As a final caveat, we highlight that given the uncertainty of X-ray preheating prescriptions \citep[e.g.][]{2014Natur.506..197F}, it is crucial to understand the ripples of using more realistic models of X-ray preheating. Having confirmed the robustness of our large-scale boxes to X-ray preheating models, we leave a more detailed exploration of the role of X-ray preheating in the effect of patchy reionization on the \lya forest to future work.  

\subsection{Caveats and limitations}
\label{ssec:limi}
Our current multi-scale implementation of the impact of the memory from inhomogeneous reionization in the \lya forest has multiple setbacks. The use of one code for the transmission of the \lya forest and another one for inhomogeneous reionization results in the loss of higher-order correlations between the ionization field on large scales and the small-scale density field.

\ion{He}{II} reionization \citep{2020arXiv200205733U} is absent in our simulations. It is a non-trivial endeavor to estimate this effect in the memory of \ion{H}{I} reionization without a full implementation of patchy \ion{He}{II} reionization in our small-scale simulations since there are two competing effects as follows. The additional photoheating in \ion{He}{III} regions would make the relaxation of the HEMD temperature-density relation faster; nevertheless, the injected energy from the reionization process will make the gas suffer less recombinations. As a result, a delay in the relaxation process would be expected. As shown in \cite{2018MNRAS.474.2173H} (see their \S 6.3), the addition of \emph{sudden} \ion{He}{II} reionization leads to a decrease of the sensitivity to the thermal relics from \ion{H}{I} reionization of approximately fifty per cent at $z = 2.5$, i.e.\ $\partial \ln T (z = 2.5)/\partial \ln T (z = 8) = 0.031$ (without Helium) vs $\partial \ln T (z = 2.5)/\partial \ln T (z =8) = 0.017$ (with Helium). Thus, our results may be off-target by roughly a factor of two in the low redshifts $z\sim 2-3$. Inclusion of \ion{He}{II} reionization is beyond the scope of this work; however, the effect of \ion{He}{II} reionization in the HEMD phase remains one of the goals for future work in the memory from inhomogeneous \ion{H}{I} reionization in the \lya forest.

Furthermore, our approach misses the effect of clustering in the ionizing sources, which tends to make the overdense regions more transmissive \citep[see, e.g.,][]{2014PhRvD..89h3010P}. As pointed out in \cite{2019MNRAS.487.1047M}, the affected scales and the strength of this effect is similar to the one studied in this work, and hence to lowest order one should add the fluctuations in the low-redshift ionizing background and the memory of inhomogeneous reionization to any near-future precision cosmology programs aimed at the \lya forest power spectrum.

Our simulations also lack the effects of AGN feedback in the \lya forest power spectrum. This effect is stronger in the small-scales and at low redshifts, which are the opposite trends of patchy reionization in the \lya forest power spectrum. \cite{2020arXiv200202822C} shows (see their Figure 4) that $\Delta P^{\rm 1D, AGN}_{\rm F}/P^{\rm 1D}_{\rm F} \, \approx -0.062$ at $k = 0.005 \ \textup{s/km}$ and $z=2$, in comparison to $\Delta P^{\rm 1D, reio}_{\rm F}/P^{\rm 1D}_{\rm F} = 0.008$ at the same $k$ and $z$ in our work. However, for higher redshift, say $z = 4.0$, the effect of AGN feedback results in the fractional difference $\Delta P^{\rm 1D, AGN}_{\rm F}/P^{\rm 1D}_{\rm F}$ in the range of $(-0.02, 0)$ at $k = 0.005 \ \textup{s/km}$, in contrast to the reionization effect $\Delta P^{\rm 1D, reio}_{\rm F}/P^{\rm 1D}_{\rm F} = 0.054$ at the same $k$. Hence both effects can dominate at different scales and redshifts of interest.

All of the aforementioned effects are relevant for the thermal relics from \ion{H}{I} reionization; however, we deem them beyond the scope of this work. In order to eventually achieve precision cosmology with the memory from inhomogeneous \ion{H}{I} reionization, all of these effects, and other systematics already accounted for in observational efforts, must be either mitigated or accounted for.

\section{Summary and Conclusions}
\label{sec:conclu}
We explore the impact of inhomogeneous reionization on the \lya forest power spectrum and its potential as a window into the cosmic reionization and cosmic dawn in this paper. For this purpose, we incorporate two different simulation codes because the dynamical range presented herein is too large with only one simulation. We use the modified {\sc Gadget2} code to resolve the gas to below the Jeans mass prior to reionization for the small-box simulations allowing us to resolve the HEMD gas, and use a modified version of {\tt 21cmFASTv1.3} to extract the effect of reionization patchiness on the matter distribution for the large-scale simulations. However, this methodology has multiple limitations, primarily in our lack of inclusion of patchy \ion{He}{II} reionization which could reduce the impact of this systematic significantly, and generate its own broadband systematic in the \lya forest, because the IGM has not relaxed into the usual temperature-density relation before the next helium cosmic reionization occurred \citep{2020arXiv200205733U}. Furthermore, our two-scale-codes-one-IGM approach misses correlations between the high resolution small-scale structure and the patchy large-scale structure. However, we deemed this an acceptable strategy because of the huge dynamical range needed for one code to be able to compute the change in the transparency of the IGM due to inhomogeneous reionization.

Using our methodology and the updated \lya forest results from eBOSS + BOSS, we computed the effect of patchy reionization on the 1D \lya forest power spectrum and found it to be larger than previously reported in \cite{2019MNRAS.487.1047M}. This difference was mainly due to the correction of a typo in the 3D to 1D integration from \cite{2019MNRAS.487.1047M}. Also, here we use more recent observational data and higher resolution in comparison to our previous work. We study the change of this power spectrum with different reionization and thermal histories and obtain the predictions for signal strengths ranging from $0.78$ per cent to $1.40$ per cent at $z_{\rm obs} = 2.0$ and from $11.0$ per cent to $18.9$ per cent at $z_{\rm obs} = 4.0$. The strength of this effect is comparable to (larger than) the observational error bars present in the measurements \citep{2019JCAP...07..017C} at low redshifts (at high redshifts). Efficient separation of the astrophysics of reionization from the 1D \lya forest power spectrum remains to be quantified in future work.  

Moreover, we computed the effect of inhomogeneous reionization on the 3D \lya forest power spectrum for different models of reionization, and found that the effect is generic, i.e. it does not vanish for any model. We found predicted changes of $3.19 - 6.24$ per cent at $z_{\rm obs} = 2.0$ and deviations of $28.5 - 51.3$ per cent at $z_{\rm obs} = 4.0$. In general, models that influence the reionization history the most produce the largest variations; however, in this work we did not holistically tackle the challenge of X-ray preheating of the IGM in our small-scale boxes. We will explore more realistic treatments of X-ray preheating and its impact in the transparency of the IGM in future work. In conclusion, due to the significance of the impact of patchy reionization in the \lya forest, steps must be taken to ensure that ongoing and near-future instruments that plan to measure the \lya forest power spectrum are not hindered by this systematic.   

More importantly, we showed the distinct behavior (with respect to wavenumber) of this effect of patchy reionization which depends on the astrophysical parameters of cosmic reionization and cosmic dawn, particularly parameters of the bubble models. As such, the \lya forest 3D power spectrum at the post-reionization epoch has the potential to distinguish between various astrophysical models by exploiting different $k$-dependence of the power spectrum. This is a promising sign for efficient separation of the astrophysical information from cosmological information, and thus opens a new window -- in principle --  into the EOR and, if further studies show larger impact due to X-ray preheating, possibly even the cosmic dawn. To achieve this, there are a few possibilities. For example, one may extract the quadrupole of the 21~cm power spectrum for the information of cross-power spectrum. Alternatively, by exploiting the fact that the effect is diminished at lower redshifts, from which cosmology may be extracted, the \lya forest power spectrum at higher redshifts may be used to constrain the astrophysics of the reionization process. In future work, we will explore the plausible scenario where \lya forest 3D power spectrum will be measured by instruments like DESI without mitigation scheme implemented, to determine what information could be extracted regarding the reionization and thermal histories.

\section*{Acknowledgements}
This work is supported by the National Key R\&D Program of China (Grant No.~2017YFB0203302, 2018YFA0404502), and the National Natural Science Foundation of China (NSFC Grant No.11761141012, 11673014, 11821303). PMC was supported by the Tsinghua Shui Mu Scholarship. YM was supported in part by the Chinese National Thousand Youth Talents Program. We thank Christopher M. Hirata, Xiao Fang, Hayato Shimabukuro and Shifan Zuo  for fruitful discussions and valuable feedback, and thank Andrei Mesinger and Jaehong Park for useful suggestions regarding  {\tt 21cmFAST}. Moreover, we thank the anonymous referee for valuable suggestions and feedback. The small-scale simulations were done in the Ruby cluster at The Ohio Supercomputer Center, 
and the large-scale simulations were ran in the Venus and Orion clusters at Tsinghua University. 

\section*{Data availability}
The data underlying this article will be shared on reasonable request to the corresponding authors.




\bibliographystyle{mnras}
\bibliography{quasi} 




\bsp	
\label{lastpage}
\end{document}